\documentclass[twocolumn]{aastex631}

\usepackage{lineno}
\shorttitle{Structures in quasar host galaxies with HSC}
\shortauthors{Nagele et al.}
\graphicspath{{./}{}}

\begin{document}

\title{A machine learning approach to assessing the presence of substructure in quasar host galaxies using the Hyper Suprime-Cam Subaru Strategic Program}

\author{Chris Nagele}
\affiliation{Department of Astronomy, School of Science, The University of Tokyo, 7-3-1 Hongo, Bunkyo, Tokyo 113-0033, Japan}

\author{John D. Silverman}
\affiliation{Kavli Institute for the Physics and Mathematics of the Universe (WPI), The University of Tokyo, Kashiwa, Chiba 277-8583, Japan}
\affiliation{Department of Astronomy, School of Science, The University of Tokyo, 7-3-1 Hongo, Bunkyo, Tokyo 113-0033, Japan}
\affiliation{Center for Data-Driven Discovery, Kavli IPMU (WPI), UTIAS, The University of Tokyo, Kashiwa, Chiba 277-8583, Japan}

\author[0000-0001-6742-8843]{Tilman Hartwig}
\affiliation{Kavli Institute for the Physics and Mathematics of the Universe (WPI), The University of Tokyo, Kashiwa, Chiba 277-8583, Japan}
\affiliation{Department of Physics, School of Science, The University of Tokyo, Bunkyo, Tokyo 113-0033, Japan}
\affiliation{Institute for Physics of Intelligence, School of Science, The University of Tokyo, Bunkyo, Tokyo 113-0033, Japan}

\author{Junyao Li}
\affiliation{CAS Key Laboratory for Research in Galaxies and Cosmology, Department of Astronomy, University of Science and Technology of China, Hefei 230026, China}
\affiliation{School of Astronomy and Space Science, University of Science and Technology of China, Hefei 230026, China}

\author{Connor Bottrell}
\affiliation{Kavli Institute for the Physics and Mathematics of the Universe (WPI), The University of Tokyo, Kashiwa, Chiba 277-8583, Japan}

\author{Xuheng Ding}
\affiliation{Kavli Institute for the Physics and Mathematics of the Universe (WPI), The University of Tokyo, Kashiwa, Chiba 277-8583, Japan}

\author[0000-0002-3531-7863]{Yoshiki Toba}
\affiliation{National Astronomical Observatory of Japan, 2-21-1 Osawa, Mitaka, Tokyo 181-8588, Japan}
\affiliation{Academia Sinica Institute of Astronomy and Astrophysics, 11F of Astronomy-Mathematics Building, AS/NTU, No.1, Section 4, Roosevelt Road, Taipei 10617, Taiwan}
\affiliation{Research Center for Space and Cosmic Evolution, Ehime University, 2-5 Bunkyo-cho, Matsuyama, Ehime 790-8577, Japan}




\begin{abstract}

The conditions under which galactic nuclear regions become active are largely unknown, although it has been hypothesized that secular processes related to galaxy morphology could play a significant role. We investigate this question using optical $i$-band images of 3096 SDSS quasars and galaxies at $0.3<z<0.6$ from the Hyper Suprime-Cam Subaru Strategic Program, which possess a unique combination of area, depth and resolution, allowing the use of residual images, after removal of the quasar and smooth galaxy model, to investigate internal structural features. We employ a variational auto-encoder which is a generative model that acts as a form of dimensionality reduction. We analyze the lower dimensional latent space in search of features which correlate with nuclear activity. We find that the latent space does separate images based on the presence of nuclear activity which appears to be associated with more pronounced components (i.e., arcs, rings and bars) as compared to a matched control sample of inactive galaxies. These results suggest the importance of secular processes, and possibly mergers (by their remnant features) in activating or sustaining black hole growth. Our study highlights the breadth of information available in ground-based imaging taken under optimal seeing conditions and having accurate characterization of the point spread function (PSF) thus demonstrating future science to come from the Rubin Observatory. 

\end{abstract}



\section{Introduction} \label{sec:intro}

Most massive galaxies host a supermassive black hole (SMBH) with mass of \hbox{$M_{BH}\gtrsim10^6$ M$_{\odot}$}. Their presence is observed from the early universe at $z\sim6-7$ \citep{mortlock2011,wu2015,banados2018,matsuoka2019,wang2021}, to the quiescent SMBH at the center of our Milky Way. Their formation relies on the episodic accretion of matter through a radiatively-efficient disk on subparsec scales, which we observe as active galactic nuclei (AGN) and luminous quasars. To keep pace with the growth of their parent galaxy, the accretion disk must be supplied by matter (i.e., gas, dust) from larger kiloparsec scales, i.e., the interstellar medium of its host galaxy. However, it is unclear how the angular momentum of rotationally-supported matter is lost thus leading to the transport of matter to the nuclear region - capable of replenishing the fuel for a massive black hole.

One viable mechanism is the merger of two (or more) massive galaxies. The initial interaction through tidal torques induces gravitational instabilities (e.g., matter asymmetries) within each galaxy that then cause gas inflow to the nuclear regions that triggers a burst of star formation and accretion onto a SMBH \citep[e.g.,][]{Mihos1996ApJ...464..641M,Hopkins2010MNRAS.407.1529H,Moreno2015MNRAS.448.1107M,Blumenthal2018MNRAS.479.3952B}. Numerous studies, both theoretical and observational, have provided substantiating results from various perspectives in terms of galaxy merger state and AGN demographics \citep[e.g.,][]{Silverman2011,Ellison2011,Mechtley2016,Goulding2018,Ellison2019MNRAS.487.2491E,Toba2022arXiv220811824T}.

Over the last decade, it has become apparent that galaxy mergers play a sub-dominant role in the growth of SMBHs, even though mergers are conducive to SMBH activity (see aforementioned studies). The majority of the host galaxies of luminous AGN are shown to be star-forming, disk-like galaxies \citep{Gabor2009,Schawinski2011,Kocevski2012,Hopkins2014} without signs of undergoing an interaction or merger. The fact is that the merger rate of galaxies is too low to build the bulk of the mass in SMBHs. There remains, however, the possibility that mergers with disproportionate mass ratios may be important.


As a result, focus has shifted to internal (i.e., secular) galactic processes that can channel material to central regions \citep[e.g.,][]{Dekel2009}. These may be due to structures (spiral arms, bars, clumps) indicative of disk instabilities and possibly minor mergers not accounted for in the aforementioned efforts. The majority of studies have looked at lower luminosity or obscured AGNs where signs of these structures on both galaxy and nuclear scales may be present and easier to discern. An early attempt by \citet{Bournaud2012} found compelling evidence for possible excess AGN activity in a sample of clumpy galaxies at $z\sim0.7$ based on an automated algorithm and visual classification to identify galaxies with prominent asymmetric features. However, a subsequent study by \citet{Trump2014} found little difference in the AGN fraction of galaxies at $z>1$ separated into those that are smooth or clumpy based on visual classification of CANDELS images. Furthermore, the roles of bars in triggering AGN activity has also been elusive \citep{Cisternas2013,Cisternas2015}. While ambitious, these initial studies may have unknown systematic effects in the identification of galaxies with structures - capable of inducing nuclear gas flows, low number statistics due to the limited survey areas, and possible diversity in potential fueling mechanisms. It could also be that the galaxy structure on smaller scales (i.e., within the central kiloparsec) is more relevant \citep{Hicks2013}.


\begin{figure*}[ht!]
\includegraphics[width=\textwidth]{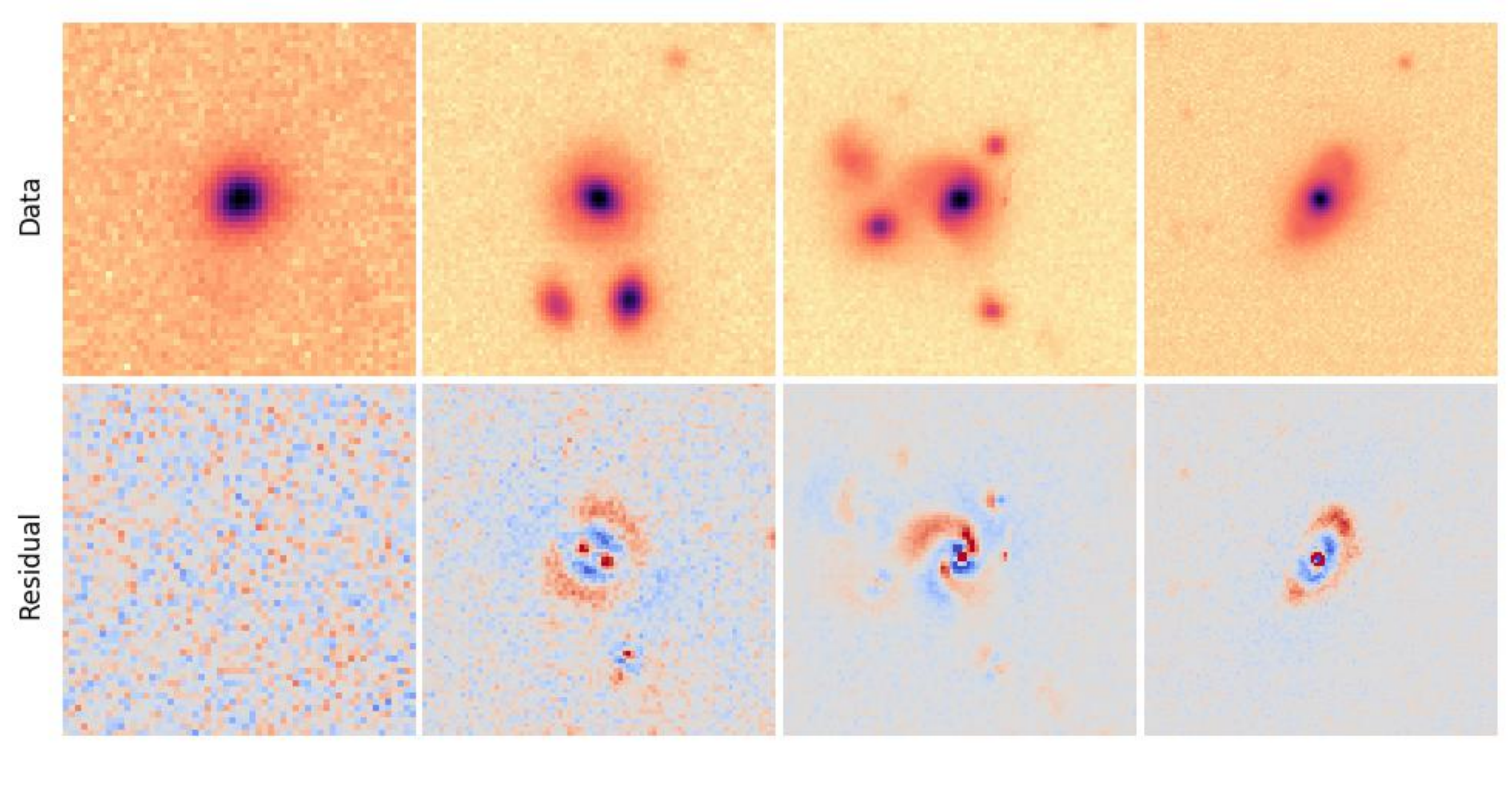}
\caption{HSC images ($i$-band) of four SDSS quasars (upper row) and their residuals (lower row; data - model quasar - model galaxy) with sub-structure seen in three cases. The residual values are generally a few percent of the signal in the original image. We have scaled each residual image to best emphasize its features.
   \label{fig:decomp}}
\end{figure*}

To date, there has been little effort on luminous quasars due to challenges of dealing with a bright unresolved point source outshining the host galaxy. Unavoidably, it is this population that is most likely to show more prominent signs of secular processes at work on galaxy-wide scales. Here, we aim to overcome the impact of the quasar light to assess the level of structure in quasar host galaxies for a large statistical sample.  We make use of a recent analysis of the host galaxies of 4887 quasars at $z  < 1$ using the five-band (grizy) optical imaging from the Hyper Suprime-Cam Subaru Strategic Survey (HSC-SSP, \citealt{Aihara2018PASJ...70S...4A}). HSC has high resolution and excellent depth \citep{Miyazaki2018PASJ...70S...1M} that enable the characterization of low surface brightness structures, which would otherwise not be detected in shallower surveys \citep{Bottrell2019MNRAS.486..390B}. In \citet{Li2021ApJ...918...22L} (L21 hereafter), the optical emission is decomposed to separate the quasar and host galaxy components using an analytic model of the point-spread function and a smooth 2D Sersic profile. As part of the data products, a residual image is produced which has both the quasar and host galaxy emission subtracted from the original science image (See Figure 2 in L21). As demonstrated, the residual emission, not accounted for with the smooth azimuthally averaged component, is easily visible in many cases with signs of bars, rings, and spiral arms. These features would not be visible at such high redshift without the depth afforded by HSC-SSP and this increased sensitivity allows us to investigate galaxy morphology in great detail.  

Characterizations of galaxy morphology \citep{Abraham1994ApJ...432...75A,Conselice2003ApJS..147....1C,Lotz2004AJ....128..163L,Pawlik2016MNRAS.456.3032P,Nevin2019ApJ...872...76N} often exhibit high sensitivity to image quality \citep{Ji2014A&A...566A..97J,Bottrell2019MNRAS.490.5390B} and to the radial extent out to which such substructures can be measured \citep{McElroy2022MNRAS.515.3406M}. In this paper, we employ a technique from generative learning--- a variational auto-encoder \citep{jimenez2014}--- that allows the compression of the residual images to a lower dimensional latent space. Other machine learning methods are often used for classification of images, but in this work we aim to gain an understanding of what kinds of morphology appear in the presence of a quasar. We accomplish this by analyzing specific areas of the latent space which show an imbalance in a set of quasar hosts and a matched galaxy sample. We then determine which kinds of substructures are associated with those areas of the latent space. 

We assume a flat cosmology with $\Omega_\Lambda$ = 0.7, $\Omega_m$ = 0.3, and $H_0 = 70\rm\ km\ s^{-1}\ Mpc^{-1}$. All magnitudes are given in the AB system. In Sec. \ref{sec:HSCdata} we describe the data, in Sec. \ref{sec:methods} we describe the variational auto-encoder, after which we report our results in Sec. \ref{sec:results} and finish with conclusions in Sec. \ref{sec:discussion}.

\bigskip \bigskip \bigskip

\section{Sample and imaging data} \label{sec:HSCdata}

\subsection{HSC optical imaging of SDSS quasars} 

For our analysis, we select a sample of 3096 broad-line (type 1) quasars at $0.3 < z < 0.6$ from the SDSS DR14 catalog \citep{Paris2018} that have five-band ($grizy$; \citealt{Kawanomoto2018}) optical imaging from the Second Data Release (DR2) of the HSC SSP survey \citep{Aihara2019} which covers about 300 deg$^2$. These represent SMBHs that are accreting at high rates as evident by their high luminosities and may show a connection with structural properties of their hosts as mentioned above. The 5$\sigma$ depth reaches $\sim26$ mag in each optical band. The upper limit on the spectroscopic redshift is chosen to ensure that the host galaxies are detected for a high fraction of the quasars (see Figure 14 of L21). The lower limit is set to restrict the width of the redshift range to minimize the variation in projected physical scale and maintain a large enough sample for this analysis. The inclusion of quasars depends on specific flags, descriptive of the quality HSC imaging, as described in L21 and \citet{Bosch2018}. The sample used here is a subset of the 4887 SDSS quasars at $z < 1$ with HSC imaging presented in L21. For our purpose here, we will only use the $i$-band images which have the highest spatial resolution due to the median seeing conditions of 0.6$^{\prime\prime}$. We refer the reader to L21 and \citet{Ishino2020} for full presentations on the study of quasar hosts with HSC.

\begin{figure*}[ht!]
\centering
\includegraphics[width=\textwidth]{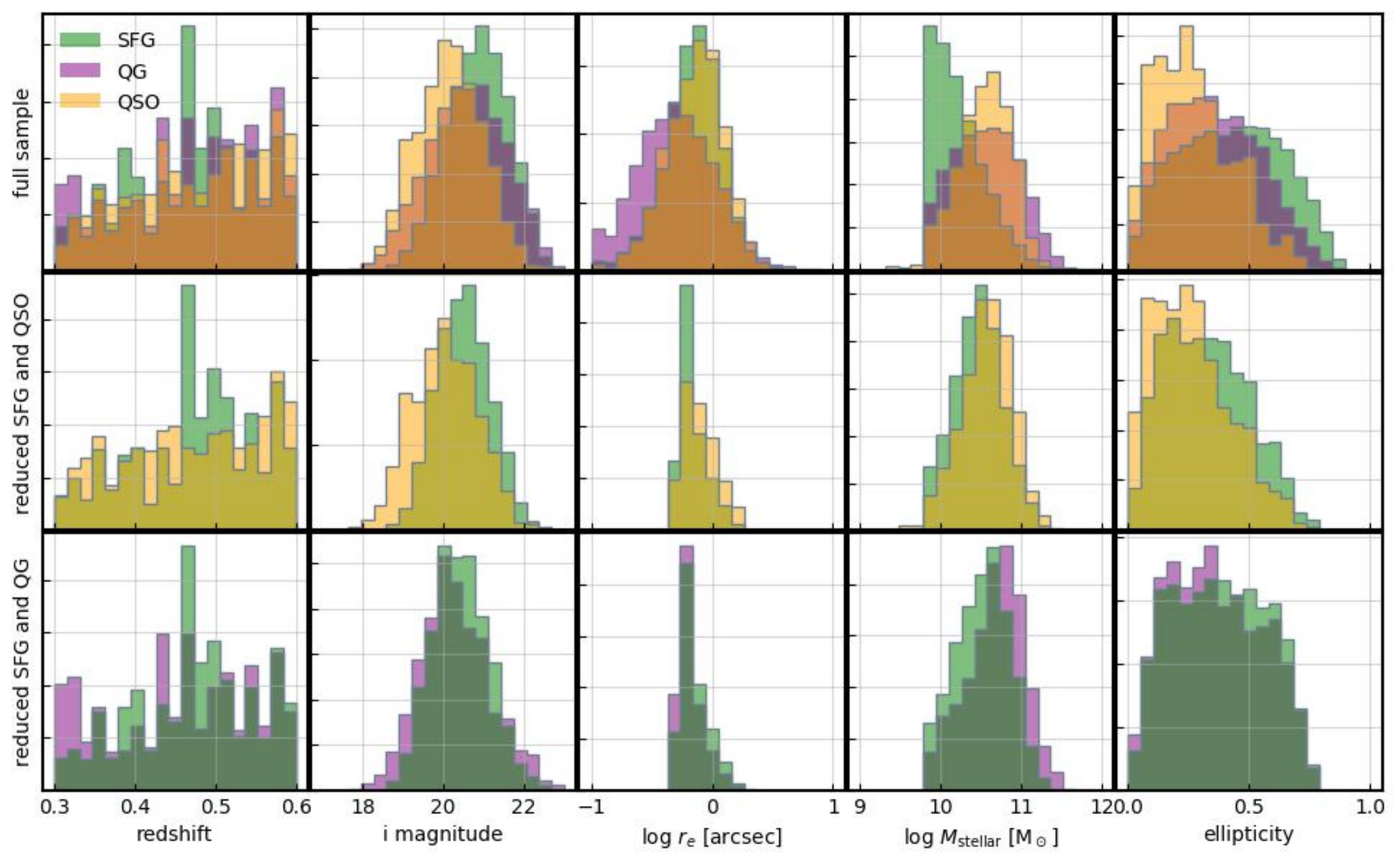}
\caption{Normalized histograms of sample characteristics. Top row --- distributions of five variables for populations of star-forming galaxies (SFG), quiescent galaxies (QG), and quasar hosts (QSO). The five variables are redshift, $i$-band magnitude, half-light radius ($r_e$), stellar mass, and ellipticity. Middle row --- same as top row, but for the sample of SFG and QSO used in the VAE. The sample is first limited to a range of $r_e$, after which we match the SFG sample to the QSO sample in mass and ellipticity (Sec.~\ref{sec:input}). Bottom row --- same as middle row, but for QG matched to SFG. \label{fig:stats}}
\end{figure*}

Stellar mass measurements of the host galaxy are based on the decomposed HSC photometry. The SED fitting code CIGALE \citep{Boquien2019} is used to determine the best-fit model SEDs which incorporates a delayed star formation history (SFH), stellar population synthesis models \citep{Bruzual2003}, a \cite{Chabrier2003} initial mass function, a \cite{Calzetti2000} attenuation law, and a nebular emission model. We set a minimum limit on the stellar mass of the host galaxy for the sample to be $10^{9.8}$ M$_{\odot}$ which results in a mass distribution mainly between $10^{10}$ and $10^{11}$ M$_{\odot}$.  In Figure ~\ref{fig:stats}, we show the distribution of the relevant properties of the sample.

\subsection{2D image analysis: quasar + host galaxy modeling}
\label{sec:2d_analysis}
We use the tool LENSTRONOMY \citep{Birrer2018} to forward model in 2D the optical emission of each quasar+host system. This analysis was prior to the availability of Galight \citep{Ding2021galight} which simplifies the process for use with large data sets. The model components include an unresolved point source to characterize the quasar emission based on an empirical model of the PSF, determined by using $\sim70$ nearby stars detected on the same CCD as the quasar \citep{Coulton2018,Carlsten2018}. The host galaxy is modeled as a single S$\acute{e}$rsic profile. The free parameters are the half-light radius $r_e$, S$\acute{e}$rsic index $n$, normalizations (host+quasar), host ellipticity (1--b/a where a and b are the semi-major and semi-minor axis respectively) and position angle. The two component model is convolved with the PSF. In the fitting, the position of the point source can be up to 0.3$^{\prime\prime}$ offset from the center of the host. The best-fit parameters are determined by LENSTRONOMY that uses a Particle Swarm Optimization  \citep{Kennedy1995} which reduces the chance of falling in a local minimum.       

We input to LENSTRONOMY a cutout science image, noise map, and empirical model PSF.  The original cut-out size is determined separately for each quasar to be between $41\times41$ pixels ($7\arcsec\times7\arcsec$) and $131\times131$ pixels ($22\arcsec\times22\arcsec$) so that the emission from the quasar and close companions are included and jointly modeled, as can be seen in the middle two panels of Figure~2 of L21.

Residual images are generated by subtracting from the science frame the best-fit model of the quasar and host galaxy. Examples of residual images are shown in Figure~\ref{fig:decomp}. By eye, there is evidence for structure in many of the residual images that is not accounted for by the smooth models. Our premise is that such features would be more common among galaxies hosting quasars either because they are driving nuclear accretion (e.g., bars) or because they share a common origin with the unknown driver of the accretion. Thus our aim is to assess whether these structures are more common or not in quasar hosts as compared to a matched non-active galaxy sample.

\begin{figure*}[ht!]
\centering
\includegraphics[width=\textwidth]{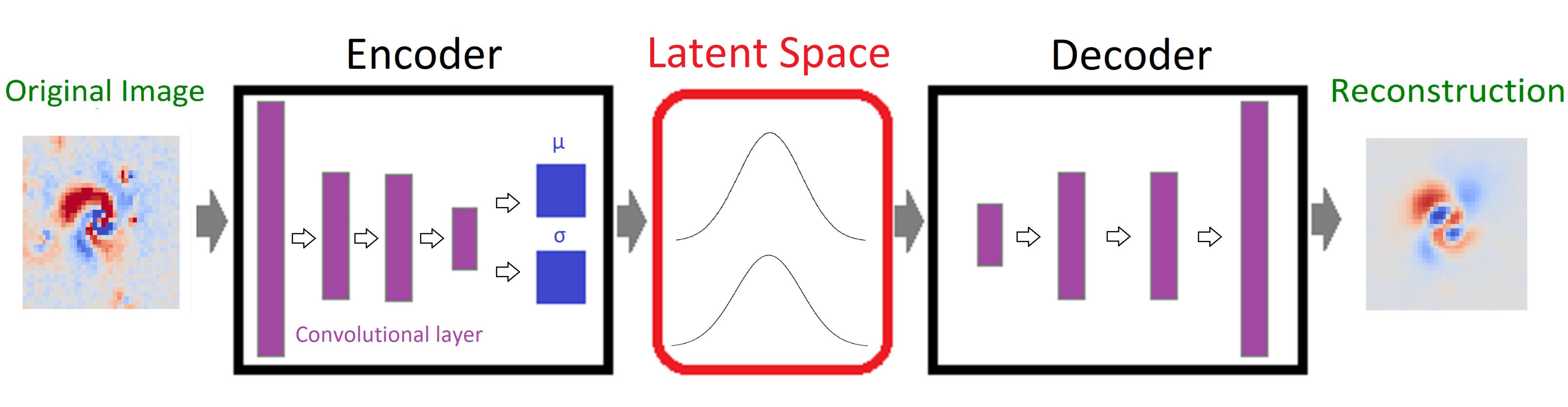}
\caption{Schematic diagram of the VAE. The original image is fed into the encoder which uses a CNN and dense layers to output a distribution ($\mu,\sigma$) in the latent space. The decoder samples from that distribution (this is the \textit{variational} element) and uses another CNN to make the reconstructed image. The weights of the various layers are initiated randomly, and then updated during training using the loss function from Eq. \ref{Eq:loss}. \label{fig:schem}}
\end{figure*}

\subsection{Inactive galaxy sample from HSC}
\label{sec:inactive}
We construct a comparison sample of 24581 galaxies from \citet[][K21 hereafter]{Kawinwanichakij2021ApJ...921...38K}. These galaxies are selected from $\sim100$ square degrees of the HSC SSP Wide survey area and do not have any considerable quasar emission. Photometric redshifts and stellar masses are computed using the Bayesian template fitting-code \textsc{Mizuki} \citep{Tanaka2015}. Any offsets between stellar masses derived by MIZUKI and CIGALE are minimal and do not impact this study. Using the same analysis tool (LENSTRONOMY), K21 has carried out single-component model fits using a S$\acute{e}$rsic profile. 

This sample is a subset of the 1.5 million galaxies with such measurements in HSC fields presented by K21. Using the best-fit parameters, we randomly draw five galaxies for each quasar with similar properties to reproduce the distributions of redshift, size (half-light radius in arcsec), stellar mass and ellipticity of the quasar host galaxies. The latter is required since quasar selection is known to result in hosts being more face-on due to extinction preventing more edge-on systems to be included in quasar catalogs (see L21). K21 has further subdivided the sample into star-forming and quiescent galaxies (SFG and QG respectively) based on their rest-frame $u-r$ vs. $r-z$ colors. The lack of infrared imaging over the full HSC Wide area survey prevents the use of the commonly used $UVJ$ method. Here we have 16739 star-forming and 7842 quiescent galaxies in this sample. We note that this classification is not perfect. There is contamination at the level of 8\% for star-forming and 12\% for quiescent galaxies (Figure 15 of K21) that should not severely impact the results of our study.     

Residual images are generated in a different manner depending if the host galaxy is star-forming or quiescent. In the latter case, a single Sersic profile for the host galaxy is sufficient to model the emission. For star-forming galaxies, a single Sersic profile may not best capture the central, essentially unresolved, emission from a bulge component. This may induce features in the residual map which are a result of the mismatch between having different number of model components between different samples which we want to avoid. Therefore, we include an unresolved point-source component to the single Sersic profile. This ensures that the model used in the fitting for SFGs exactly matches the two-component model used for the quasars and their hosts. This applies to the first two experiments listed in Table~\ref{tab:experiments} as described further below. 

For one experiment, we compare the results for star-forming and quiescent galaxies without quasars in each sample. In this case, a single Sersic profile is used in the modeling of both inactive galaxy samples. This is the third experiment listed in Table~\ref{tab:experiments}.

\subsection{Simulated quasar host galaxy sample}
\label{sec:sim-qso}
It is expected that some of the features seen in the residual images shown in Figure~\ref{fig:decomp} are artifacts due to an imprecise empirical model of the PSF. The reasons can be attributed to spatial variations of the PSF and a dependence on color since the color of the stars used to construct the PSF have not been matched to that of the quasar which is difficult due to the lack of stars having similar colors on each CCD.

To assess the impact of the PSF, we have generated a set of residual images of a simulated quasar sample using exactly the same decomposition routine as described above. The simulated quasar sample is constructed by adding unresolved PSF models to real HSC images of SFGs (see Section~\ref{sec:inactive} for classifying HSC galaxies). The model unresolved quasar emission has Poisson noise added which is characteristic of the HSC images. Specificially, we use a model PSF at the position of the galaxy to add a simulated quasar component. Then, for the decomposition, we use a different model PSF, randomly chosen from an offset position from the galaxy center by $\pm1^{\prime\prime}$ in either the E-W or N-S direction. The simulated quasar sample spans the same range in parameters as the observed SDSS/HSC quasars in terms of redshift, stellar mass, host-to-total flux ratio, and host galaxy size and ellipticity. The procedure to simulate quasars and their hosts is similar to that employed in L21.

\begin{deluxetable*}{lllllll}
\tablenum{1}
\tablecaption{Descriptions, parameters, and results of the three experiments. \label{tab:experiments}}
\tablewidth{0pt}
\tablehead{
\colhead{}&\colhead{Purpose}&\colhead{Samples\tablenotemark{a}} & \colhead{\# of images\tablenotemark{c}} & \colhead{min$\{P_{max}(F_{200}(z)) \}$\tablenotemark{d}} & KS distance\tablenotemark{d} & KS p\tablenotemark{d}
}
\startdata
$\#1$&Test-PSF removal&SFG, SIM-QSO\tablenotemark{b} &  1387 , 1387 & 4.60e-20 & 0.051 & 3.30e-13  \\
$\#2$&Experiment &SFG, QSO &  1387 , 483 & 2.50e-40 & 0.296 & 9.69e-223 \\
$\#3$&Experiment&SIM-QSO, QSO & 1387 , 483 & 2.37e-33 & 0.325 & 2.15e-269\\
$\#4$&Experiment&SFG, QG & 1826 , 1136 & 4.31e-42 & 0.430 & $<$1e-300\\
\enddata
\tablenotetext{a}{SFG: star-forming galaxy, QSO: quasi-stellar object, SIM-QSO: simulated QSO (Sec.~\ref{sec:HSCdata}), QG: quiescent galaxy}
\tablenotetext{b}{SIM-QSO: simulated quasars+hosts using the galaxies in the SFG sample.}
\tablenotetext{c}{The number of images is after the reductions in Sec.~\ref{sec:input} are made, and the number of images which go into the VAE are this number times the augmentation factor.}
\tablenotetext{d}{The final three columns are described in Sec.~\ref{sec:latent}}

\end{deluxetable*}

\section{Methods} \label{sec:methods}

We are now well within the era of data-driven astronomy. Since telescopes produce gigabytes of data per second, we need efficient algorithms to digest and analyse this data in order to address previously untenable questions. In particular, for the complex task of galaxy classification, the community has relied on visual inspection \citep[][]{Nair2010ApJS..186..427N,trump14,Sola2022A&A...662A.124S} including large-scale citizen science projects \citep{GalaxyZoo}. Recently, however, tools such as convolutional neural networks (CNNs, \citealt{leCun1989,lecun1998,Dominguez2018MNRAS.476.3661D,Bottrell2019MNRAS.490.5390B,Ciprijanovic2020A&C....3200390C,Cheng2021MNRAS.507.4425C,Bickley2021MNRAS.504..372B,Walmsley2021AAS...23811902W}) and variational auto-encoders (VAEs, \citealt{kingma2013,jimenez2014,nishikawa-toomey2020,spindler2021}) have been utilized for analyzing large imaging data sets. These AI-based methods can be trained to obtain a high-level, abstract (i.e., lower dimensional) representation of input images that can then be used for classification or morphological analysis.

\begin{deluxetable}{cc}
\tablenum{2}
\tablecaption{Hyper parameters of the VAE (see Sec.~\ref{sec:opt}).  \label{tab:params}}
\tablewidth{0pt}
\tablehead{
\colhead{Hyper-parameter} & \colhead{Value}
}
\startdata
        Learning Rate & $10^{-3}$ \\ 
        Batch Size & 128 \\ 
        Reconstruction Loss Factor, $\lambda$ & 2.5 $\times$ $10^4$ \\ 
        Latent Space Dimension, $z$  & 18 \\ 
        Sigmas Clipped, $m$ & 5 
\enddata

\end{deluxetable}

\subsection{Variational Auto-encoder} \label{sec:VAE}

Generative methods are machine learning methods which involve the model generating mock data as part of the training process \citep{foster2019}. They have been successfully applied to several astronomical problems \citep{ravanbakhsh2016,Schawinski2017MNRAS.467L.110S,sun2019,formsma2020,cai2020,Portillo2020AJ....160...45P,arcelin2021,Boone2021AJ....162..275B,Villar2021ApJS..255...24V,Hemmati2022ApJ...941..141H} including the investigation of galaxy morphology using VAEs \citep{nishikawa-toomey2020,spindler2021}, though the morphology of quasar hosts has not yet been studied. VAEs are a type of dimensionality reduction, an approach which has seen widespread use in the characterization of galaxy observations \citep[e.g.][]{Vanderplas2009AJ....138.1365V,Rahmani2018MNRAS.478.4416R,Hemmati2019ApJ...881L..14H,Davidzon2022A&A...665A..34D,Cooray2022arXiv221005862C}. VAEs boast a wide range of applicability, but are particularly useful when used together with CNNs for image analysis \citep[e.g.][]{gregor2015}.

For a computer, an image is just a two dimensional array of pixel values. In order for a computer to perform a useful analysis on a set of images, we need to incentivize the computer to digest the images into a meaningful representation that reflects, e.g., the presence of spiral arms, clumps, bars, asymmetries, etc. We achieve this with two computational procedures. First, we use convolutional layers, which preserve the correlation of nearby pixels and allow the computer to obtain a more abstract representation of the original images by identifying features, for instance the presence of edges or gradients. Second, we compress a high dimensional image ($40^2$ pixels) into a point in a lower dimensional latent space, and furthermore we incentivise the set of images to be evenly distributed throughout this latent space. Thus, we can assess the similarity of the input images by comparing their positions in the latent space, with nearby images being similar and more distant images being different.

The VAE simultaneously accomplishes these two procedures using a bottleneck architecture as illustrated in Figure~\ref{fig:schem}. An input image is first compressed by the encoder, a series of four convolutional layers with a final fully connected layer. The output of the encoder is a Gaussian distribution in the latent space which is then randomly sampled to produce a point in the latent space, which we identify with the input image. Note that the provision of a randomly sampled Gaussian distribution is the \textit{variational} component of the VAE. Then, the latent space point is fed into the decoder, consisting of four deconvolutional layers, which outputs a reconstruction image. The VAE then compares the input image to the reconstruction image and decides how well it has done, i.e. how much the reconstruction looks like the original. Based on this evaluation, the VAE improves its weights and the process is repeated.

More formally, a VAE can be seen as a variational Bayesian approach to inferring a numerically intractable posterior distribution using a neural network \citep{kingma2013,jimenez2014}. Specifically, if a data-set $d$ has a probability distribution $p$ over a set of variables $z$, then computing $p$ requires the integration over d$z$ which is difficult in general. The VAE instead computes an inferred distribution $q$. By assuming a specific form for $p$ (a standard Gaussian) and the number of variables determining the distribution (a hyper parameter of our study, see Sec.~\ref{sec:opt}), $q$ may be computed with machine learning techniques using a loss function of the form \citep[][]{foster2019}:

\begin{equation}
    \rm{Loss} =  \rm{Loss_{KL}} +\lambda  \; \rm{Loss_{RMS}},
    \label{Eq:loss}
\end{equation}
where $\lambda$ is a hyper parameter. The first term is the Kullback-Leibler divergence \citep{kullback1951} which incentivizes the inferred distribution $q$ to approximate a unit Gaussian:
\begin{equation}
    \rm{Loss_{KL}}= \frac{1}{2} \sum_{k=1}^K \bigg( \mu_k^2 + \sigma_k^2 - \log \sigma^2_k -1    \bigg),
\end{equation}
where $K$ is the dimension of the latent space. The second term is the reconstruction loss which incentivizes $q$ to approximate $p$. For this term, we use the root mean square, which is a common choice for image analysis:

\begin{equation}
    \rm{Loss_{RMS}} = \sqrt{\frac{\sum_1^N (d_n - d_n')^2}{N}},
\end{equation}
where $N$ is the number of pixels in an image, $d_n$ is the pixel value in the original image and $d_n'$ is the corresponding pixel value in the reconstruction (see below).

\begin{figure}[ht!]
\centering
\includegraphics[width=0.5\textwidth]{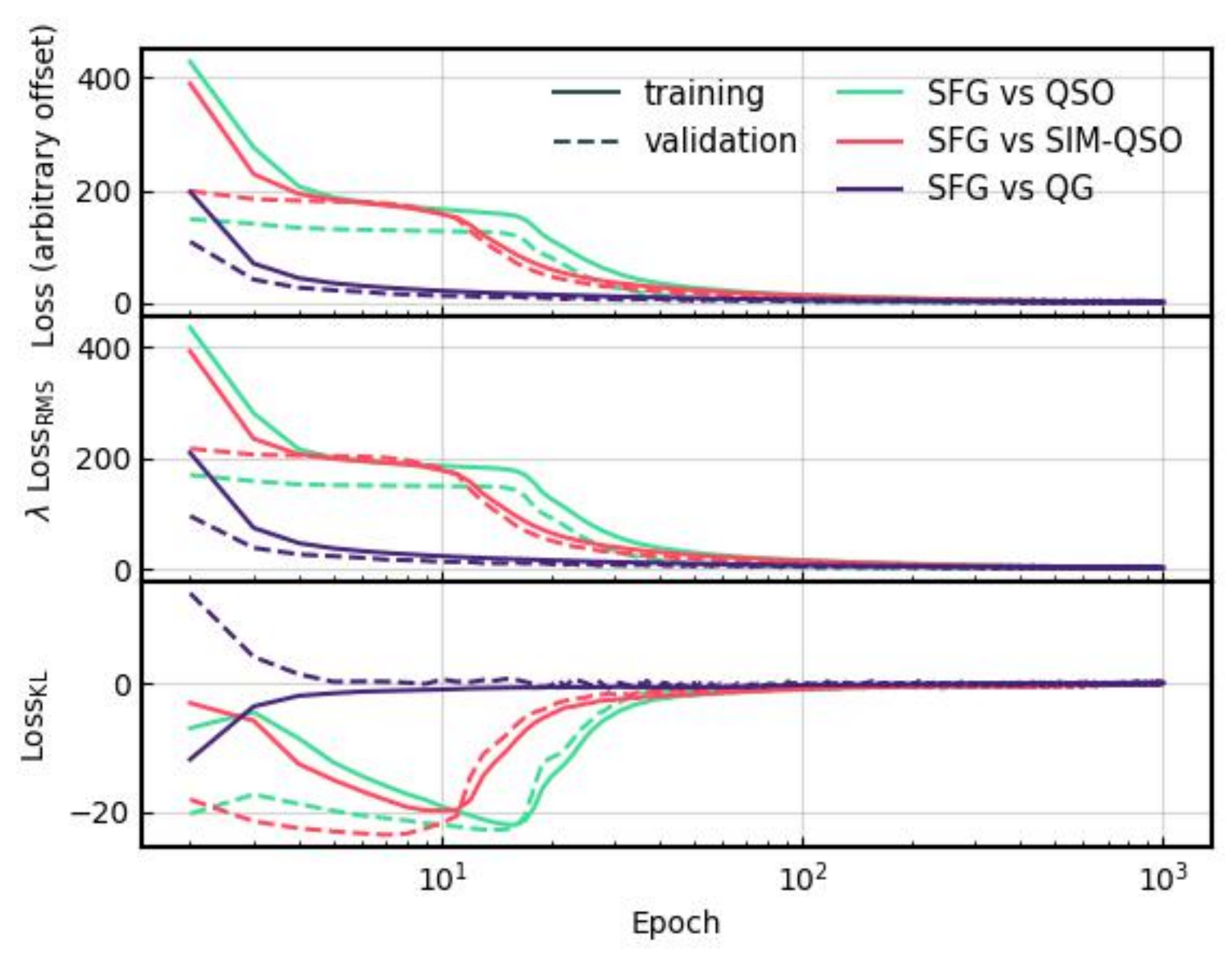}
\caption{Decomposition of the total loss (upper panel) into the reconstruction loss (middle panel) and the KL divergence loss (lower panel) for each of the three experiments, where each loss has been offset to end at zero. The validation loss is shown by the dashed lines.   \label{fig:loss_val}}
\end{figure}

\begin{figure*}[ht!]
\centering
\includegraphics[width=\textwidth]{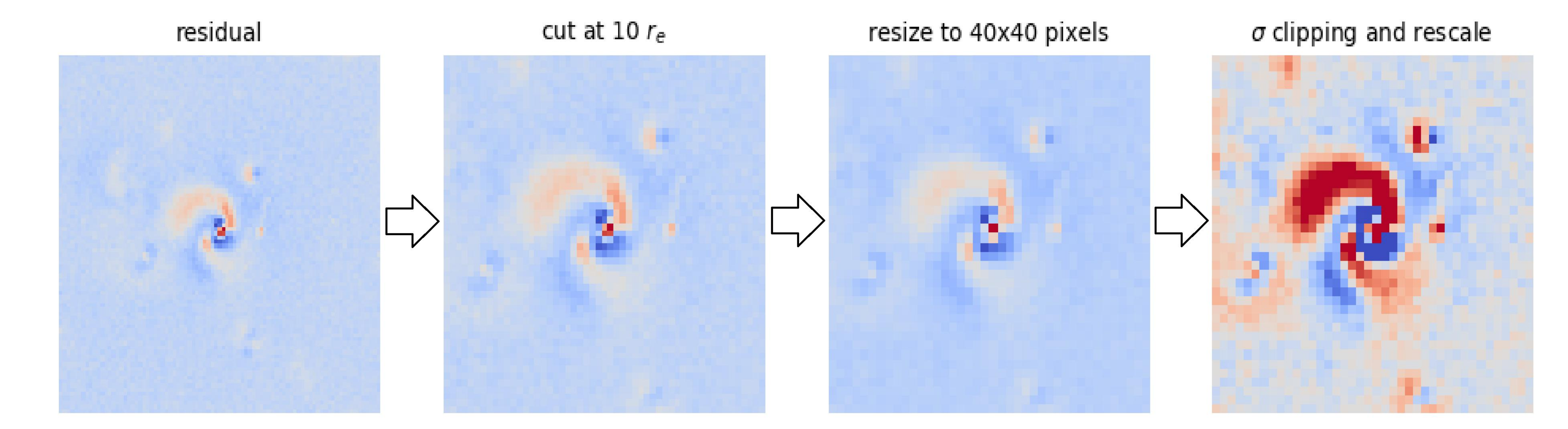}
\caption{Image preparation for the VAE. The first two steps satisfy the requirements that all images should be the same dimensions (40$^2$) and effective scale ($10\times r_e$). We then clip outlying pixels and rescale the image to the interval 0 to 1. \label{fig:vae_prep}}
\end{figure*}

Fig.~\ref{fig:loss_val} shows the loss of the training set compared to the validation set (upper panel) and the decomposition of the training loss into the reconstruction loss and the Kullback–Leibler (KL) divergence term, (middle and lower panels respectively). In each case, the total loss decreases (as expected), but at some point the VAE begins to increase the KL divergence term, effectively exchanging the gaussianity of the latent space for an improvement in the quality of the reconstructions. The degree to which this compromise occurs is determined by the hyper-parameter $\lambda$ (Sec.~\ref{sec:opt}).

The convolutional layers in the encoder have strides (2, 2, 1, 5) and kernels (3, 3, 3, 6) while the decoder has strides (5, 1, 2, 2) and kernels (6, 3, 3, 3). All layers use Leaky RELU activation functions. We use batch normalization and dropout layers \citep[e.g.][]{foster2019}.

\subsection{Preparation of the input images for the VAE} \label{sec:input}

We desire that the VAE should group images based on morphology, such that images with similar morphological features appear nearby each other in the latent space; but such an outcome is far from guaranteed, as the VAE is adept at grouping images based on statistics related to pixel values or astronomical properties (such as redshift). In an attempt to overcome this difficulty we established a procedure for the preparation of the images for the VAE as illustrated in Figure~\ref{fig:vae_prep}. 

First, we scale all images so that they are on the same effective scale (side length of 20 $r_e$, where $r_e$ is the half-light radius of the quasar host galaxy). We discard any images not large enough to cover this length. Next, we resize the images to $40\times40$ in pixels, because all input images to the VAE need to have the same dimensions. The resizing is done using Lanczos interpolation \citep{lanczos50}, which is a common method thought to be more accurate than comparable methods (e.g., bicubic interpolation). We further discard any images where the dimensions would be too small (i.e., less than 30 pixels per side).

For the next step, pixel values in each image are scaled to values between 0 and 1. This procedure is non-trivial because of extended tails in the distribution of pixel flux which vary widely between targets. Typical pixel values are of order $0.1$, but a small number of pixels have values in excess of 100, likely due to cosmic rays. We scale the images from $(\bar{x}\pm m \sigma)$ to (0, 1) where $\bar{x}$ is the mean pixel value of the image, $\sigma$ is the standard deviation of the image, $m$ is an optimized hyper-parameter (Table \ref{tab:params}).

Finally, in order to increase the number of images on which we train the VAE, we augment our sample by applying various transformations. That is, for each original image, we add to the sample copies of that image which have been rotated, or mirrored. We use the original image as well as seven augmented images. The degree to which we augment the sample is constrained by computational cost.

\begin{figure}[ht!]
\centering
\includegraphics[width=0.5\textwidth]{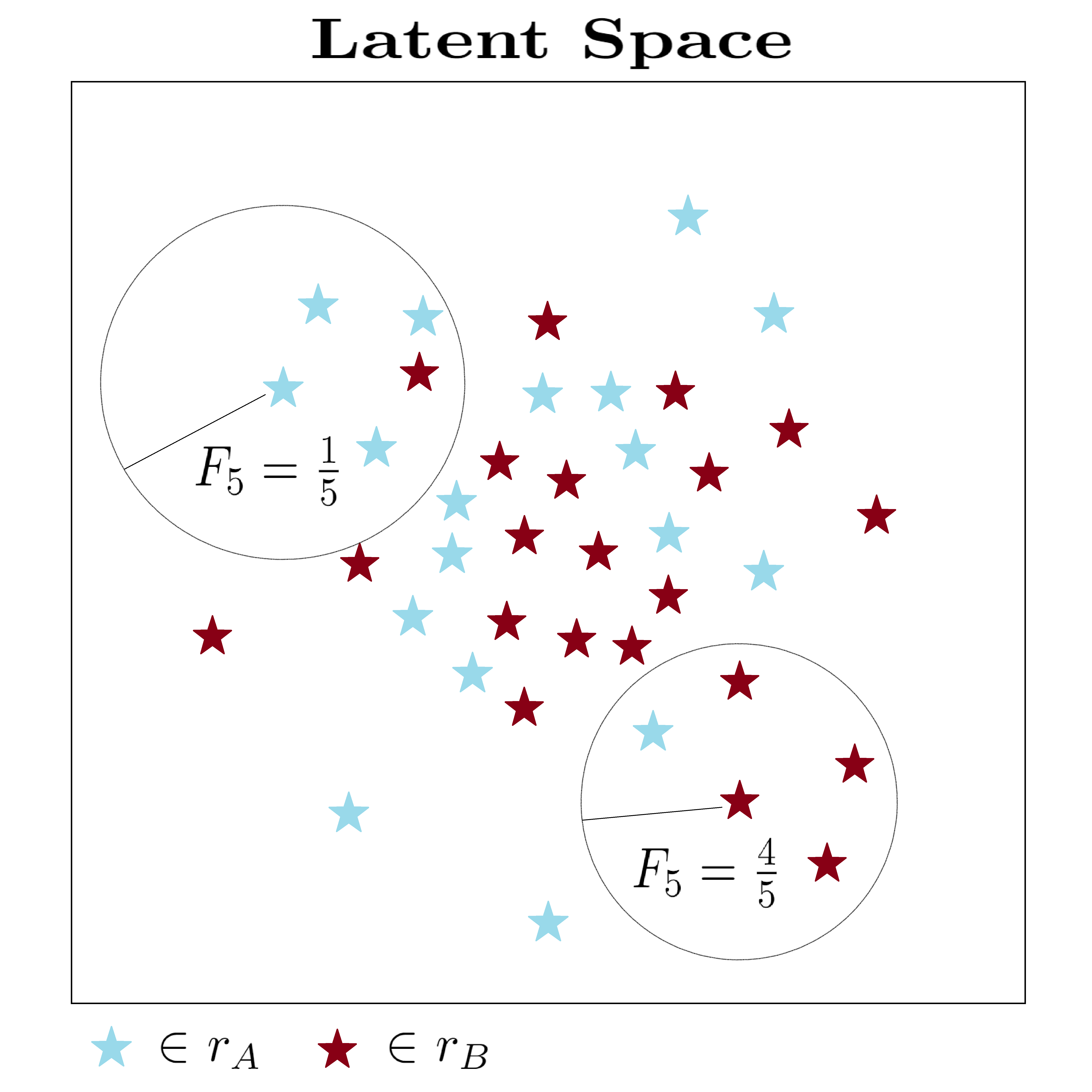}
\caption{Schematic showing the evaluation of the regional fraction belonging to one population (Eq. \ref{eq:f}). The VAE produces a latent space distribution for two populations, shown by blue stars and maroon stars. For each point in the latent space, we determine the fraction of its $x=5$ nearest neighbors which are maroon stars. \label{fig:nn_schem}}
\end{figure}

\begin{figure*}[ht!]
\centering
\includegraphics[width=\textwidth]{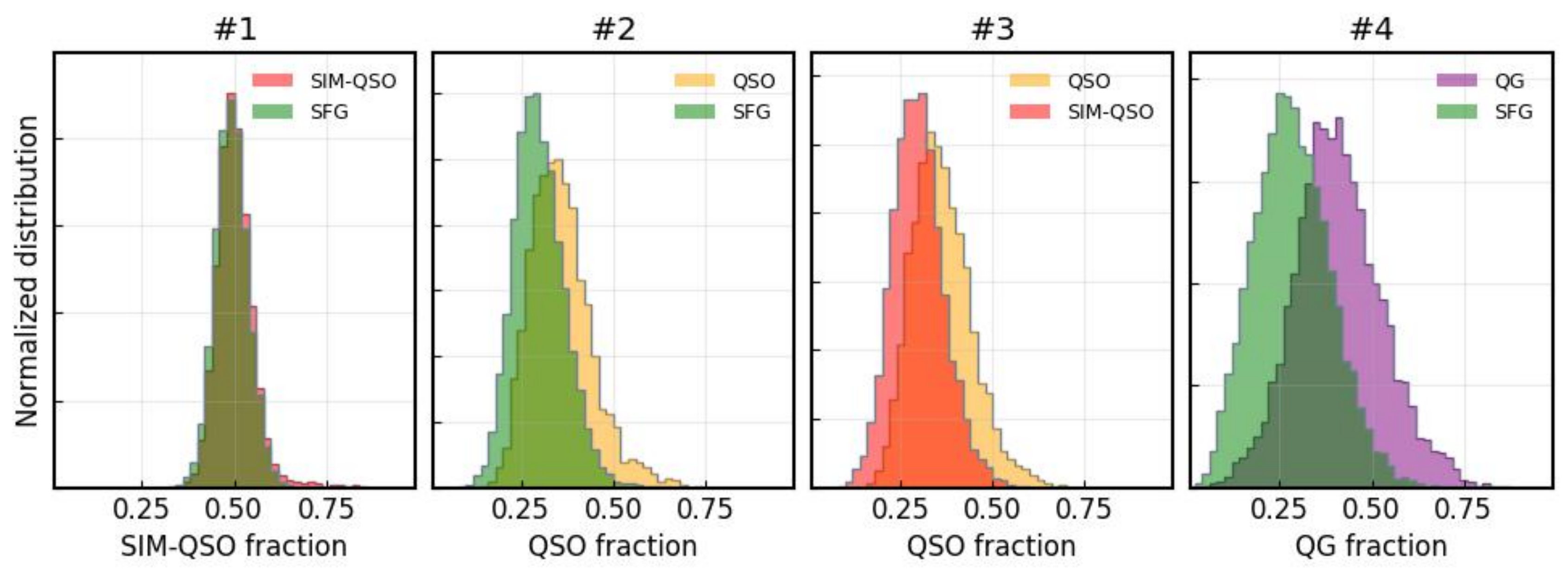}
\caption{Normalized histograms showing the distribution of population fraction ($F_{200}$) for cases as indicated in each panel: SFGs and SIM-QSOs ($\#1$), SFGs and QSOs ($\#2$), SIM-QSOs and QSOs ($\#3$), and SFGs and QGs ($\#4$). The vertical lines show the fraction of each population for the entire data-set. The results of the KS test between the two distributions are given in Table \ref{tab:experiments}.\label{fig:KS}}
\end{figure*}

\subsection{Analysis of the latent space} \label{sec:latent}
The latent space has a lower number of dimensions than that of the input image (40$^2$) and this is the primary advantage of this method. In the latent space, each original image is mapped to a point, such that images with similar features get mapped to nearby locations. We then analyze the latent space distributions by determining if there are regions where a certain population is more prevalent. Once we identify such an area, we can then determine which morphological properties are associated with that region by visual inspection.

\subsubsection{Nearest neighbor fraction}

We run the VAE on input images drawn from two data-sets (Table \ref{tab:experiments}) which we refer to here as A and B with latent space distributions $r_A$ and $r_B$ (note, $q = r_A + r_B$, but the $r$s are not themselves approximating Gaussians). The VAE does not know which images belong to each data-set, so if there were no differences (morphological or statistical) between the two data-sets, then the VAE would randomly distribute them throughout the latent space. Our primary method of analyzing the latent space is to examine the nearest neighbors of each image (Fig.~\ref{fig:nn_schem}). Specifically, we sort the images according to the distance of their latent space locations to the location of a specific encoder output $r_i$. Then, we determine the fraction of the $x$ nearest neighbors to $r_i$ which are, without loss of generality, members of data-set B:
\begin{equation}
    \label{eq:f}
    F_x(r_i) = \rm \: \: B \: fraction \:  among \; x \; nearest \; neighbors \; of \; r_i.
    \label{Eq:nn}
\end{equation}
With the above equation, we evaluate morphological differences both globally and locally.

\subsubsection{Global morphological differences}

First, we would like to know is there an overall difference between where the VAE is placing A images and B images in the latent space. To do this, we compute Eq. \ref{Eq:nn} for each image from set A, $F_x(r_A)$, and each image from set B, $F_x(r_B)$. We then compare these distributions with a Kolmogorov–Smirnov (KS) test, (Fig.~\ref{fig:KS}). If the result of the KS test is that the distributions are similar, then we conclude that the VAE does not find a difference between A and B. In the opposite case, for instance, if $F_x(r_B) > F_x(r_A)$ on average, then the VAE is not distributing the two populations evenly. This means that there is some difference between the images in A and those in B, though at this point we do not know the exact nature of the difference. For instance, it could be a single morphological difference, or it could be several morphological differences. After all, there is no guarantee that the A and B populations contain monolithic morphological features.

\subsubsection{Local morphological differences}

Another interesting question is what are the properties of the images which cluster most strongly with others from their own population. To evaluate this, we look at the extreme values of the distributions of Eq. \ref{Eq:nn}, for instance $\rm{max} \{F_x(r_B)\}$. A natural question is are the extreme values statistically significant compared to a random distribution of $r_A$ and $r_B$? 

To address this question, we calculate the probability of seeing a specific $F_x$ in the set of all possible nearest neighbor subsets, assuming a randomly distributed latent space. We use the binomial expansion to calculate the probability of one nearest neighbor subset of size $x$ having the fraction $F_x$:
\begin{equation}
    P(F_x) = \frac{{x\choose x_{B}}{n_{tot}-x\choose n_{B}- x_{B}}}{{n_{tot}\choose n_{B}}},
\end{equation}
where $x_{B}$ is the number of B images of the the x nearest neighbors and $n_{tot},n_{A},n_{B}$ are the total number of images, the number of images in set A, and the number of images in set B respectively. 

The above expression gives the probability of finding a certain fraction in a single nearest neighbor subset, assuming a randomly distributed latent space. The total number of distinct nearest neighbor subsets can vary with the latent space distribution, so rather than determine this number each time, we simply say that the number of distinct nearest neighbor subsets has a maximum of $n_{tot}$. Then, summing the probability $P(F_x)$, $n_{tot}$ times will give the upper limit on the probability of observing the largest $F_x$:
\begin{equation}
    \nonumber
    P_0 = 0
\end{equation}
\begin{equation}
    \nonumber
    P_{i+1} = P_{i} + P_F - P_{i}  P_F
\end{equation}
\begin{equation}
    P_{\rm max} = P_{n_{tot}}.
\end{equation}
If $P_{\rm max}$ for a given $F_x$ is small, then, the probability of observing that $F_x$ for a random distribution of $r_A, r_B$ in the latent space is also small. Thus the minimum of $P_{\rm max}$, occurring at one of the extreme points of the distributions in e.g. Fig.~\ref{fig:KS}, will indicate how far from random the distributions are. This quantity is recorded in Table \ref{tab:experiments}, column 3. Note that we set $x = 200$ throughout this paper. This value is selected because the KS statistic is stable for $x > 100$. There are quantitative differences in min$\{P_{max}(F_{x}(z)) \}$ for different choices of $x$, but the qualitative behaviour does not change.

\begin{figure}[ht!]
\centering
\includegraphics[width=.5\textwidth]{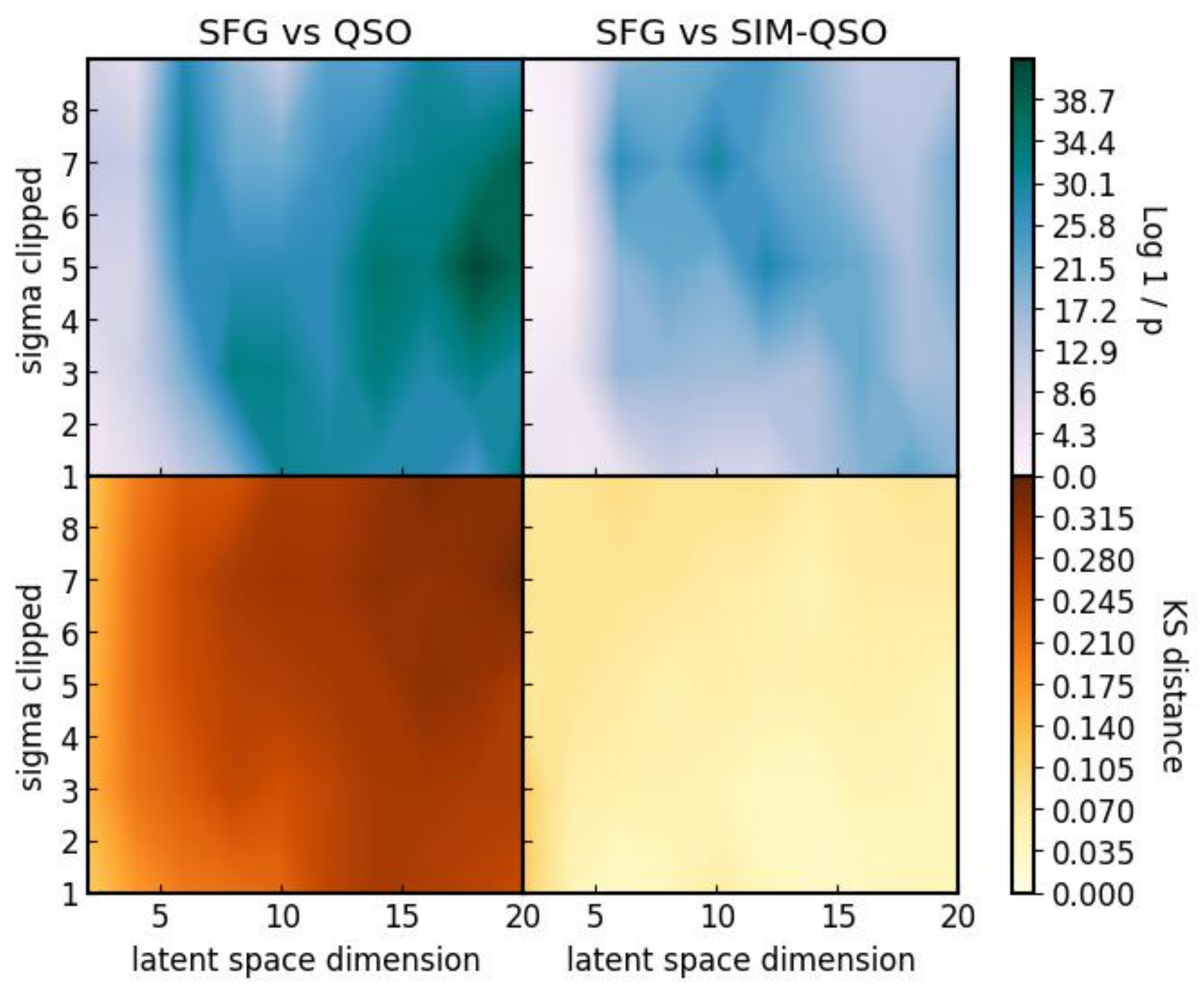}
\caption{Maps of the probability ratio (upper panels) and the KS distance (lower panels) for SFG vs QSO and SFG vs SIM-QSO (Table \ref{tab:experiments}) as a function of latent space dimension and sigma clipping. \label{fig:opt}}
\end{figure}

\subsection{VAE hyper-parameter optimization} \label{sec:opt}

The human-defined hyper-parameters of the VAE (so called to differentiate them from trainable `parameters' of the VAE) are summarized in Table \ref{tab:params}. The values of some of these parameters follow from straightforward considerations, such as the learning rate and batch size, which should be minimized and maximized respectively as allowed by computational constraints. The reconstruction loss factor ($\lambda$, Eq. \ref{Eq:loss}) is set according to a visual inspection of the reconstruction images. $\lambda$ is perhaps the most important hyper-parameter, because it determines the amount of clustering in the latent space (or more precisely, the extent to which the latent space is prevented from clustering), and for that very reason we cannot use the optimization methods described below for $\lambda$. Finding a method of optimizing $\lambda$ would be a natural next step in our investigation of the VAE. 

Two hyper-parameters which we can optimize over are the multiple $m$ for the $\sigma$ clipping (Sec.~\ref{sec:input}) and the latent space dimension $z$ (Sec.~\ref{sec:latent}). Fig.~\ref{fig:opt} shows (min$\{P_{max}(F_{200}(z)) \})^{-1}$ (top row) and the KS distance (bottom row) as a function of $m$ and $z$ for the first two experiments in Table \ref{tab:experiments}. In both rows, our goal is to find large differences between SFG and QSO, but small differences between SFG and SIM-QSO. In the top row, we can see that the lighter regions to the left of the panels (low latent space dimension) and towards the bottom of the panels (clipping a larger portion of the pixels) are both unfavorable for finding improbable latent space distributions. In the bottom row, we see that larger sigmas clipped and larger latent space dimension both allow the VAE to find larger general differences between SFG and QSO. Furthermore, SFG and SIM-QSO are similar (small KS distance) for all values of the parameters. We select $z=18$ and $m=5$.

\begin{figure*}[ht!]
\centering
\includegraphics[width=0.9\textwidth,height=0.45\textwidth]{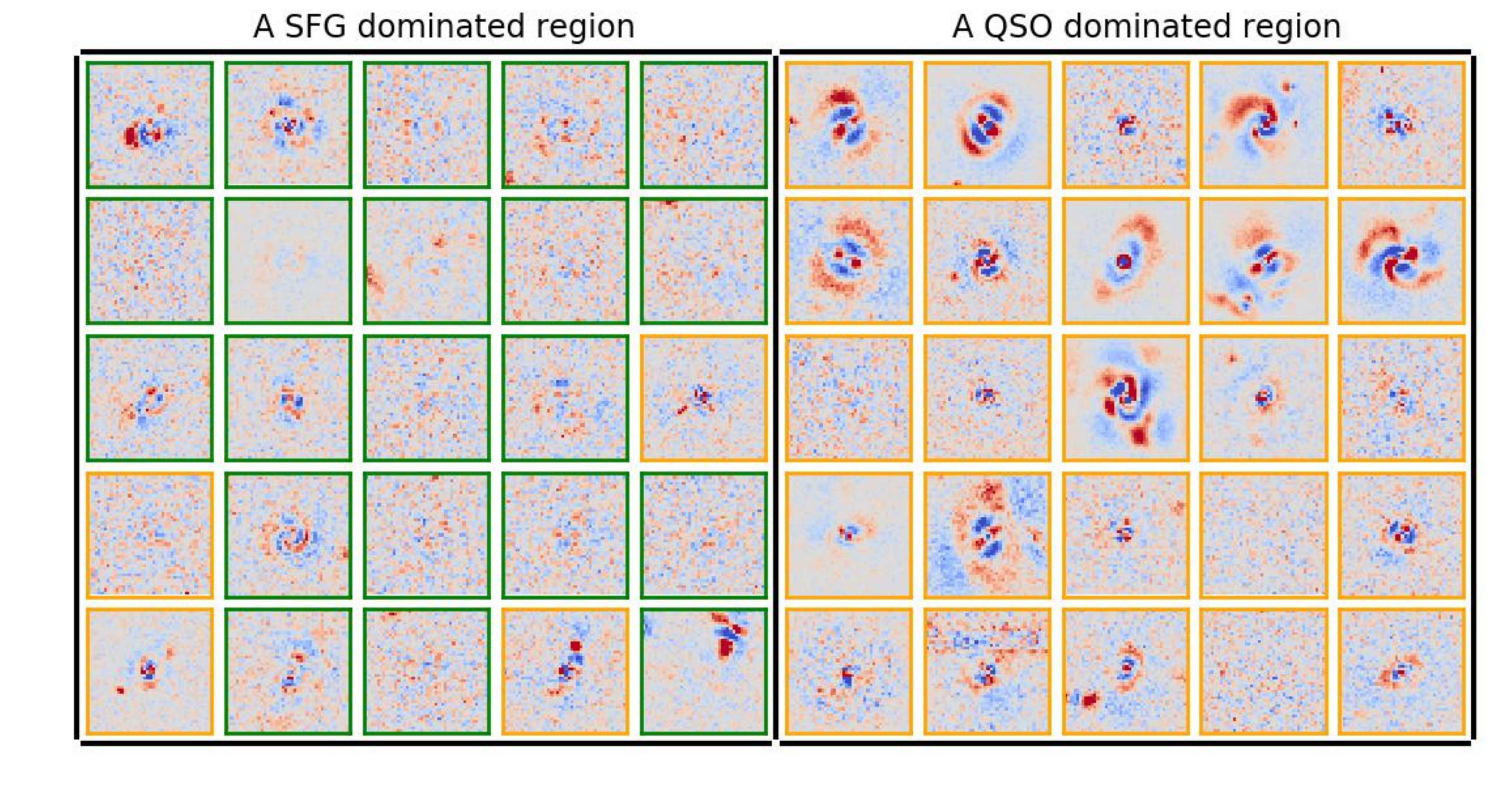}
\caption{ Gallery showing residual images from a region of the latent space dominated by SFG (left) and QSO (right). SFG images have green borders while QSO images have orange borders. The QSO dominated region shows evidence of spiral arms and rings on galaxy wide scales. \label{fig:gal_QSO}}
\end{figure*}

\begin{figure*}[ht!]
\centering
\includegraphics[width=0.9\textwidth,height=0.45\textwidth]{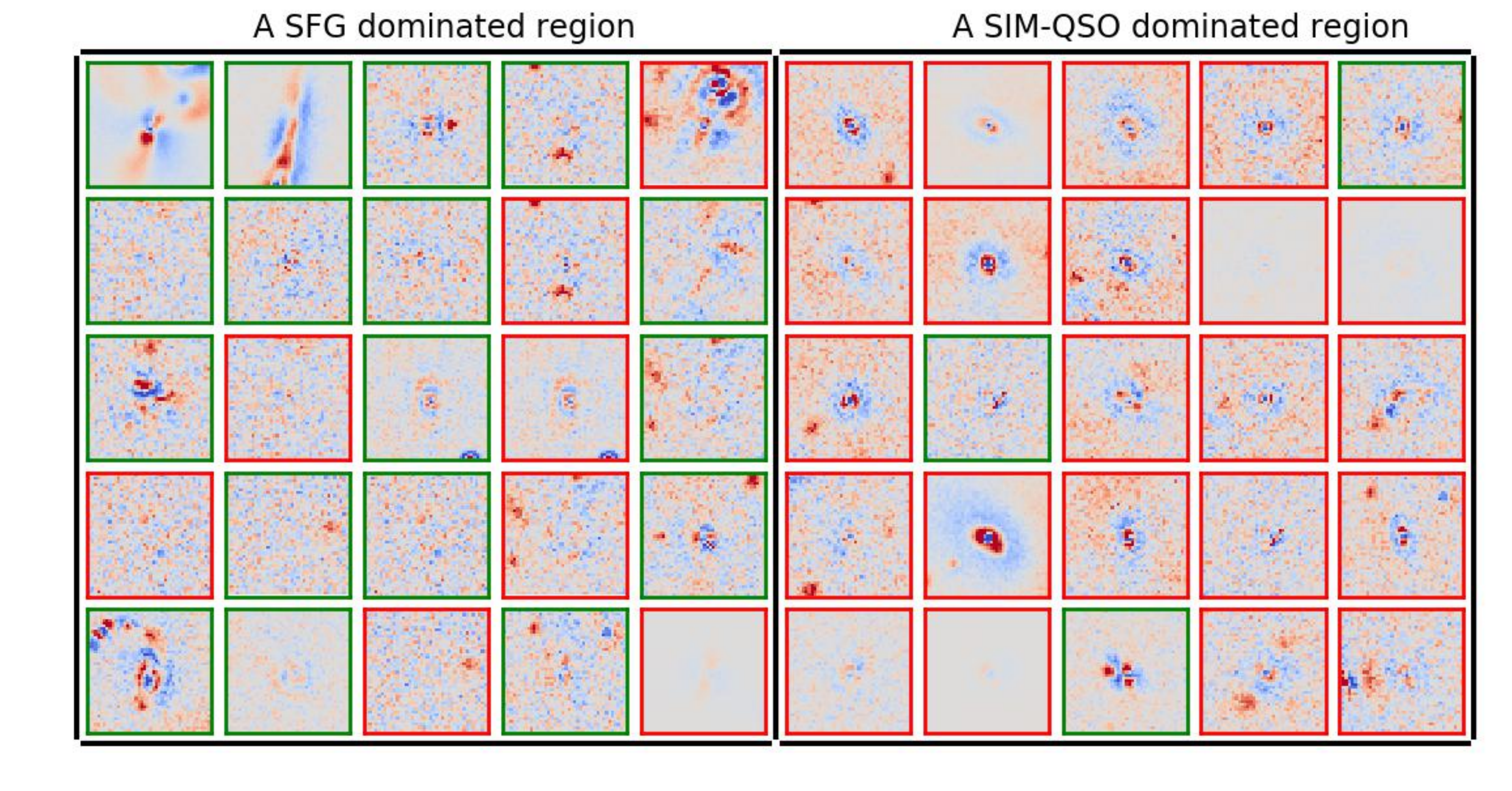}
\caption{ Same as Fig.~\ref{fig:gal_QSO} but for SIM-QSO (red borders) instead of QSO. The SIM-QSO dominated region shows a characteristic blue-red-blue ring pattern in the very center of the image.   \label{fig:gal_SIMQSO}}
\end{figure*}

\begin{figure*}[ht!]
\centering
\includegraphics[width=\textwidth]{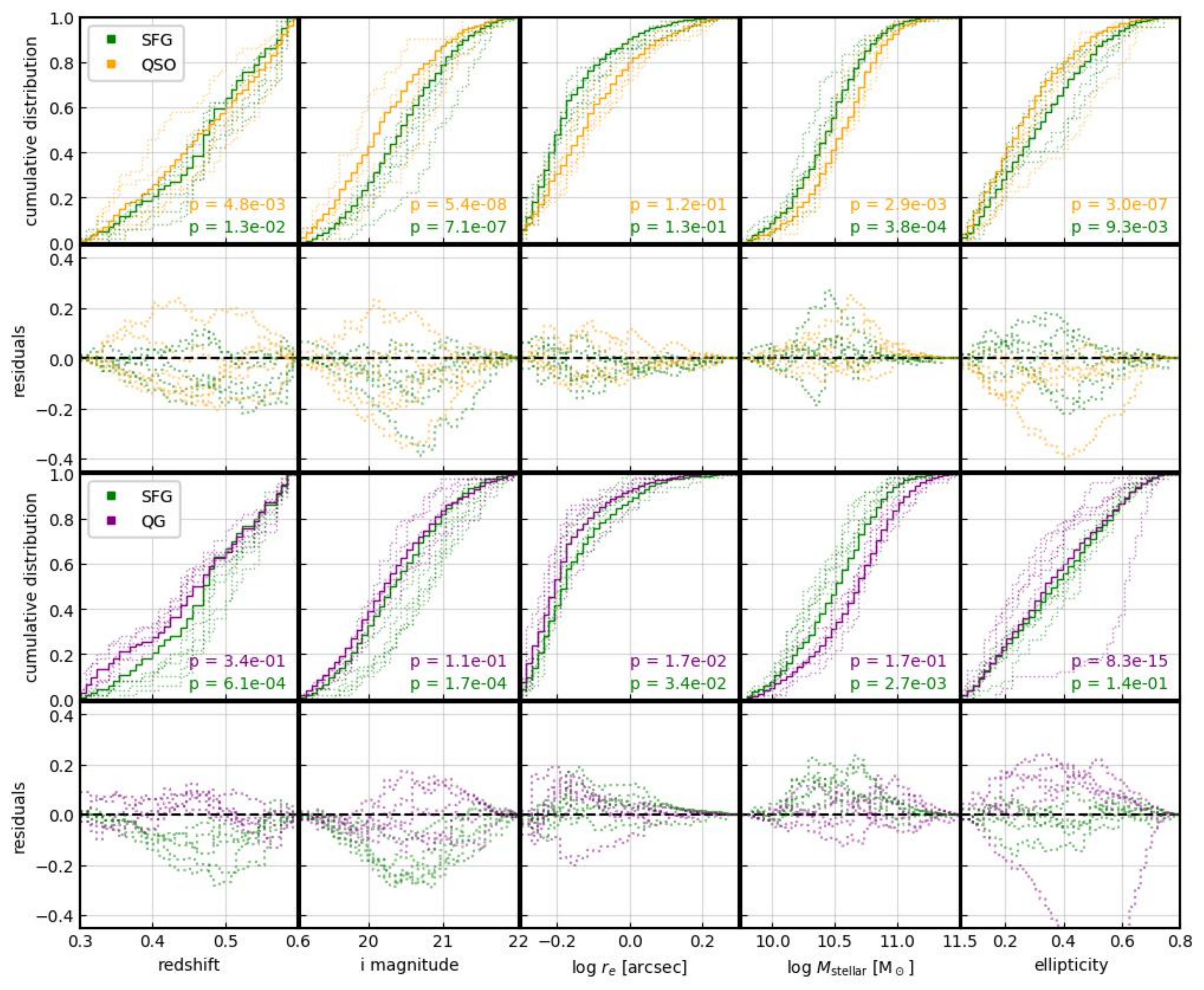}
\caption{Cumulative distributions of statistical quantities (Fig.~\ref{fig:stats}) for each population (solid lines) and for images in outlying regions of the VAE (dotted lines). The lowest KS probability from comparing the reduced sample to the outlying distributions is also shown. First row --- cumulative distributions for SFG (green) vs QSO (orange). Second row --- residuals of the first row (dotted line minus solid line). Third row --- cumulative distributions for SFG (green) vs QG (purple). Fourth row --- residuals of the third row. \label{fig:check}}
\end{figure*}

\section{Results} \label{sec:results}

We have trained the VAE on several data-sets (Table \ref{tab:experiments}) and we will now analyze the latent space for each of these experiments. This analysis is carried out primarily using $F_{200}$ (Sec. \ref{sec:latent}, Fig. \ref{fig:nn_schem}) which is the fraction of latent space points from one of the two populations among a single image's two hundred nearest neighbors in the latent space.

\begin{figure*}[ht!]
\centering
\includegraphics[width=0.9\textwidth,height=0.45\textwidth]{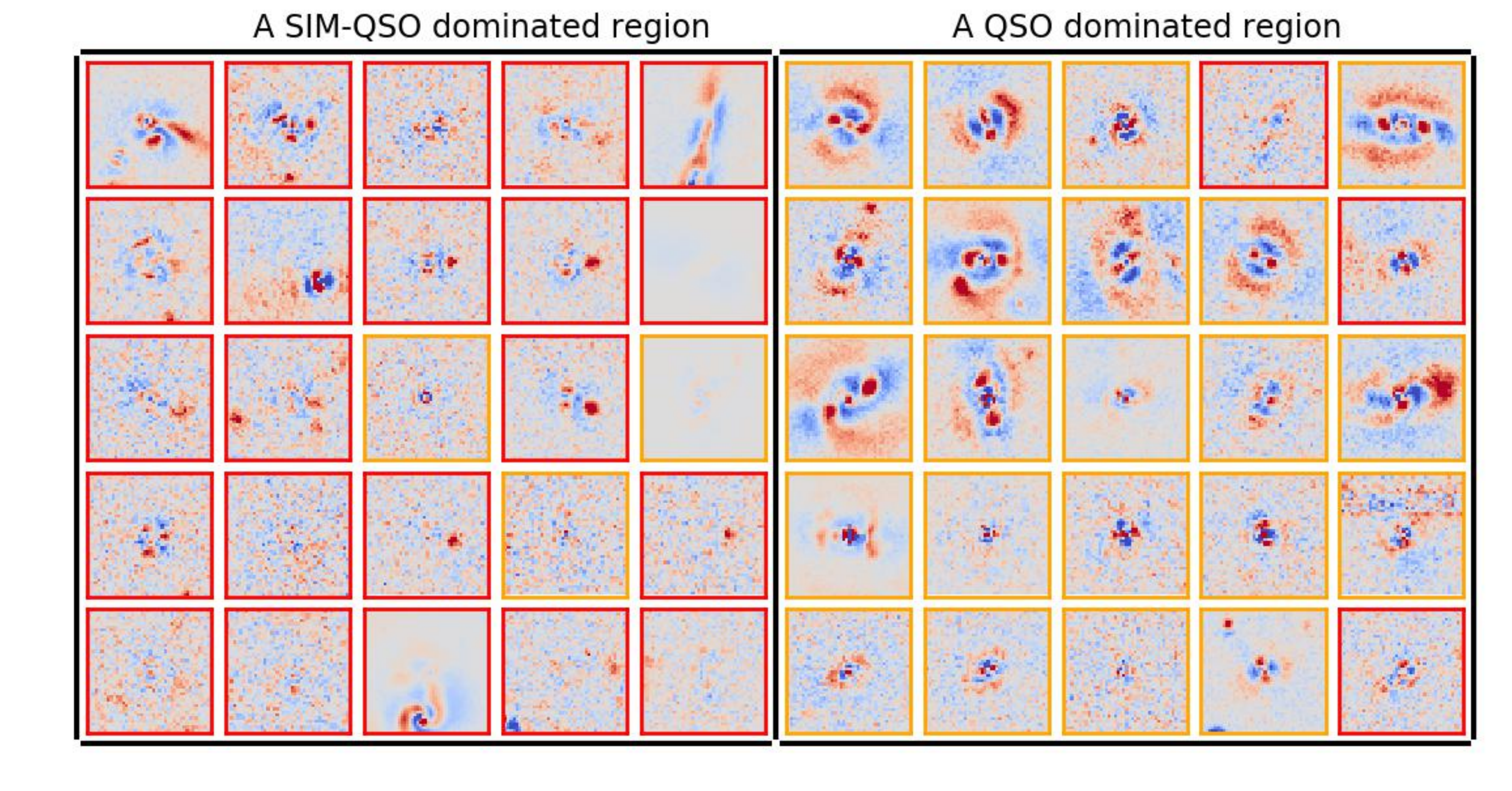}
\caption{ Same as Fig.~\ref{fig:gal_QSO} but for SIM-QSO (red borders) with QSO (orange borders). As in Fig. \ref{fig:gal_QSO}, the QSO region shows obvious rings and spirals.    \label{fig:SIMQSO-QSO}}
\end{figure*}

\subsection{Quasar hosts and star-forming galaxies} \label{sec:SFAGN}

Our first, and most scientifically interesting, experiment is to compare SFGs to QSO host galaxies. We find a large difference between the total population of SFGs and QSOs (Fig.~\ref{fig:KS}, panel $\#2$) with a KS distance of 0.296. We also find that there exist outlying regions which are dominated either by SFGs or QSOs, and the probability of finding such regions in a randomly distributed latent space is extremely small ($\sim 10^{-40}$). The most QSO dominated region (Fig. \ref{fig:gal_QSO}, right panel) shows evidence of spiral arms and rings on galaxy-wide scales (up to $\sim 5\times r_e$). Other QSO dominated regions (Appendix A; (Fig. \ref{fig:c11})) show evidence of even larger scale features, possibly remnants from past merger events.

In order to give a sense of how common these outlying regions are, we calculate the percentage of the QSO and SFG populations which have $P_{\rm max}<1\%$ ($1\%$ was chosen arbitrarily, but the results are not sensitive to this choice). This condition is satisfied for $29\%$ of the QSO sample and $16\%$ of the SFG sample. The fact that a significant portion of these populations can be classified as outliers in this way further demonstrates that the VAE preferentially groups images from the same populations together.

Subtracting the quasar signal in order to create the residuals (Sec. \ref{sec:HSCdata}) is a challenging process which may leave artifacts in some of the residual images, and the VAE could then exploit the presence of those artifacts to group images. In order to confirm that this is not driving the results discussed above, we created a sample of SFGs which then had a simulated quasar added to them before they underwent image reduction (SIM-QSOs; see Sec.~\ref{sec:sim-qso}). Thus, any artifact introduced during the quasar subtraction process should be clearly visible when comparing SFGs to SIM-QSOs. 

We find that in general, the VAE cannot tell the difference between SFGs and SIM-QSOs, with a KS distance of only 0.064 between the two populations (Fig. \ref{fig:KS}, panel $\#1$). While this is extremely encouraging, there does exist a sub-population of SIM-QSOs which were not well fit during the image subtraction process, and the VAE clusters these images together (Fig. \ref{fig:gal_SIMQSO}, right panel). Compared to the rest of the data-set, these images have on average slightly lower redshift and lower ellipticity, but there are no other clear trends. The visual signal of this artifact is a blue-red-blue ring pattern in the central area of the image ($<5r_e$). 

It is possible that the VAE is also using this artifact to group images in the SFG and QSO experiment because the probability that the outlying region of the SIM-QSO experiment comes from a randomly distributed latent space is small ($\sim 10^{-23}$); however, the probability for the QSO experiment is significantly smaller ($\sim 10^{-40}$). This fact, combined with the obvious galaxy wide morphological properties in Fig. \ref{fig:gal_QSO}, is convincing evidence that the VAE is not relying on the subtraction artifact and is in fact identifying real morphology which can then be correlated to AGN activity.

\begin{figure*}[ht!]
\centering
\includegraphics[width=0.49\textwidth]{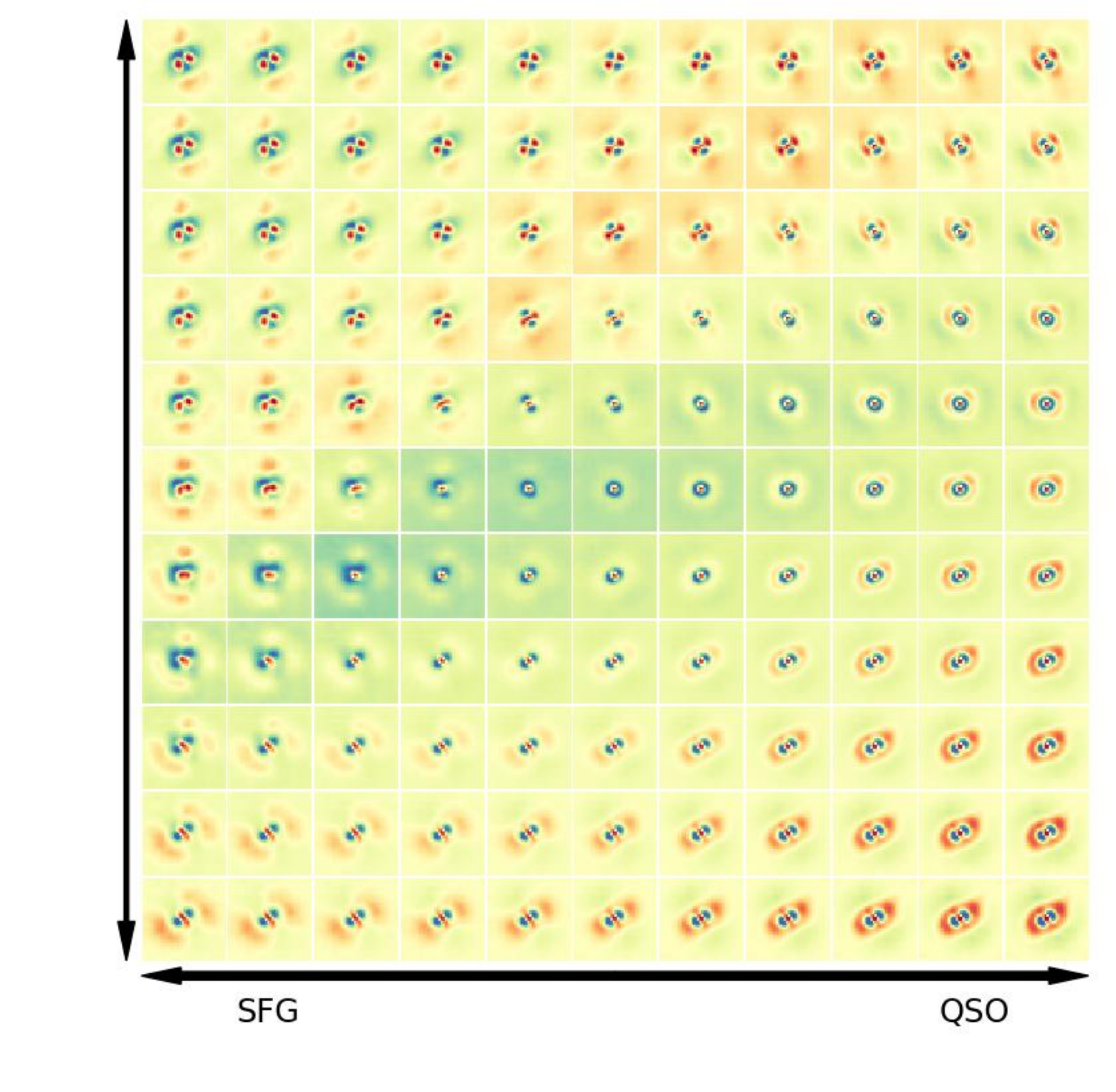}
\includegraphics[width=0.49\textwidth]{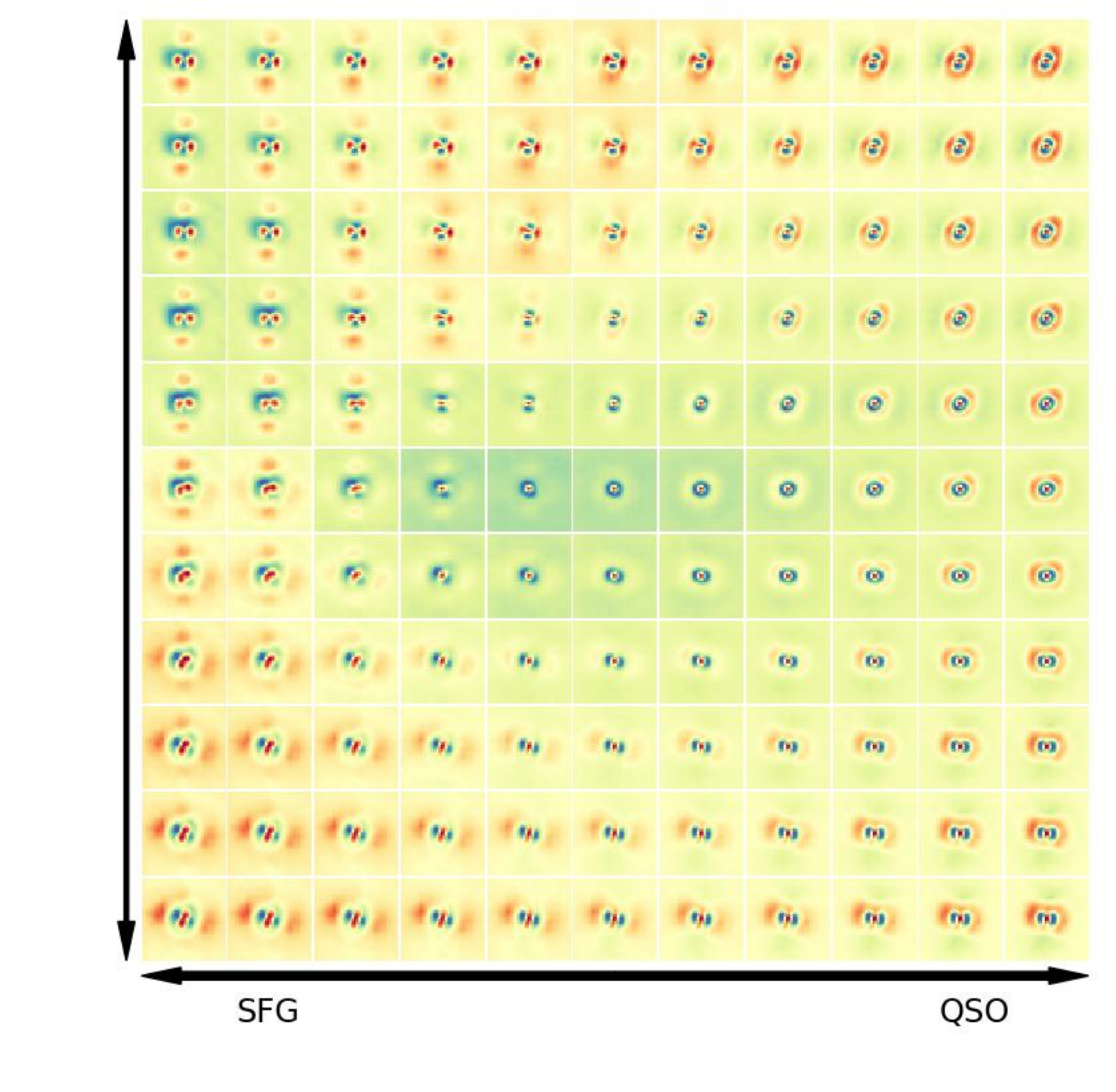}
\caption{ Two examples of latent space visualizations using reconstruction images. The horizontal axes show the change from the average latent space position of SFGs to that of QSOs while the vertical axes show two different random directions. In both panels, the central image is the average position of the entire sample.  \label{fig:re2d_map}}
\end{figure*}

\begin{figure*}[ht!]
\centering
\includegraphics[width=0.9\textwidth,height=0.45\textwidth]{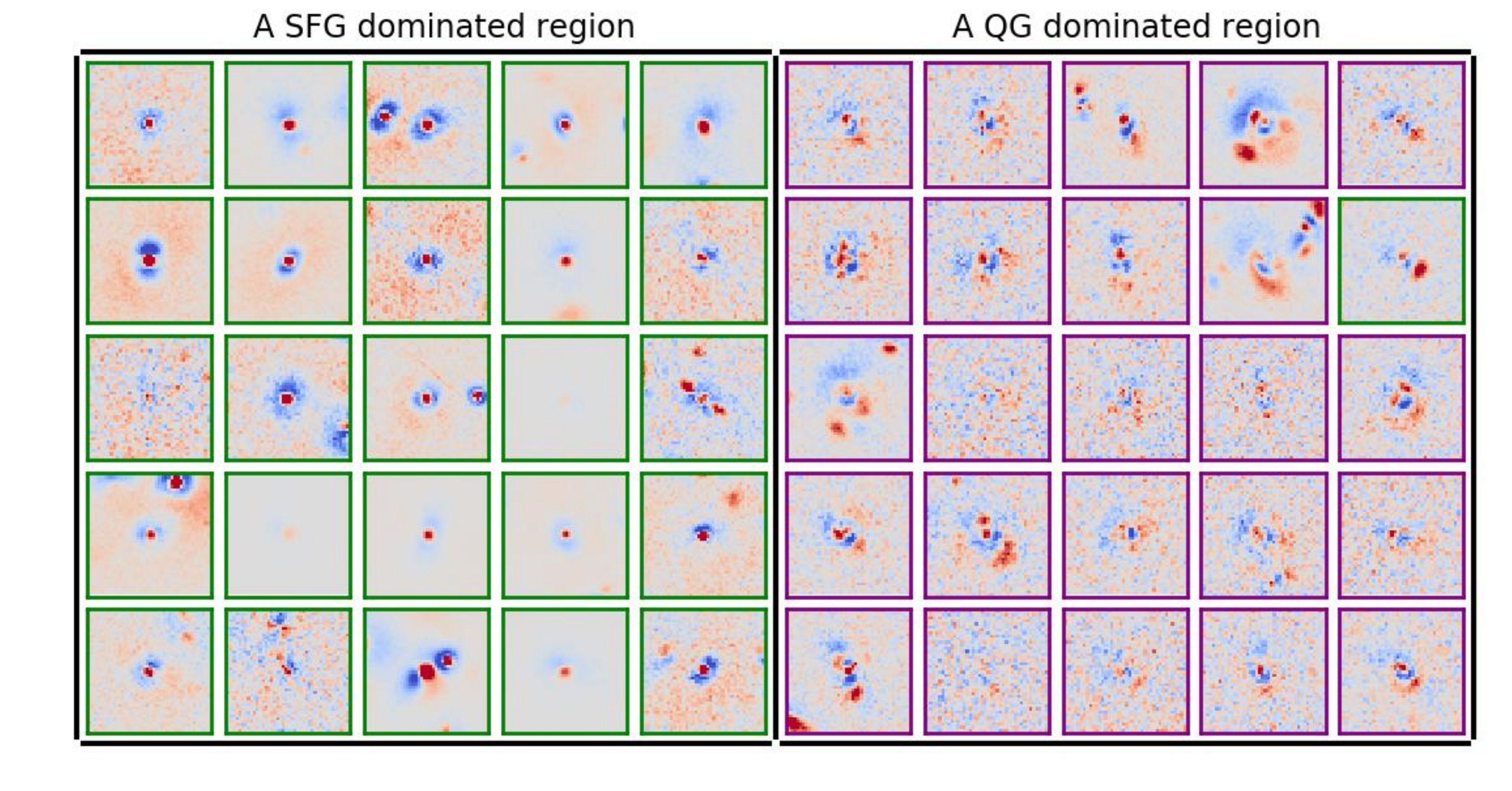}
\caption{ Same as Fig.~\ref{fig:gal_QSO} but for QG (purple borders) instead of QSO. The SFG dominated region shows a central, strong red component. The QG region has less obvious properties, but we can immediately see that it prefers asymmetric structure.  \label{fig:gal_QG}}
\end{figure*}

The caveat, of course, is that when using machine learning methods such as the VAE, it is challenging to be completely certain \textit{why} we see a certain set of results. Although the VAE was designed to group images based on morphological features, we know from experience that the VAE attempts to group images based on statistical quantities such as mass or magnitude (Fig.~\ref{fig:stats}). The first row of Fig.~\ref{fig:check} shows the cumulative distributions of the statistical quantities in Fig.~\ref{fig:stats} for both the total populations of SFGs and QSOs (solid lines) and for the sub-populations (50 images) contained in the five most outlying regions which do not overlap each other (dotted lines). The second row shows the residual of the first row (dotted line - solid line), and the third and fourth rows show the equivalent data for the SFGs and QGs.

It is clear that the outlying regions found by the VAE can correlate with some of these statistics, most extremely the QG dominated region with high ellipticity which is shown in Appendix A (Fig. \ref{fig:c06}). This behaviour is expected, because images with similar morphological properties will sometimes have similar masses, ellipcities, etc. The question that we seek to answer, then, is do the different outlying regions show similar shifts in the residuals? Indeed, for the case of SFGs and QGs, they do. The five outlying SFG regions all have lower redshift, higher $i$-band magnitude, and lower mass compared to the five outlying QG regions. Except for the outlier discussed above, this is also true of ellipticity. This behavior is not completely unexpected, as we know that morphology evolves over cosmic time and varies with stellar mass, however, it is slightly concerning because it may mean the VAE is grouping images based on these statistics instead of on morphology. 

However, this concern is not present for the main experiment, comparing SFGs to QSOs. There are some outlying regions which correlate with e.g. $i$-band magnitude, but there is no systematic difference between the statistical residuals of the SFG outlying regions and the QSO outlying regions. Thus we conclude that the VAE has not found some clever statistical shortcut when it groups images from the same population together.

As a final check, we run the VAE with SIM-QSO and QSO, which is the same as SFG and QSO where the SFGs have had a quasar artificially inserted before the image reduction process. If the subtraction artifact were driving our results, than this test would show less robust statistics than SFG and QSO, but in fact it shows even more convincing visual evidence for spirals or rings (Fig. \ref{fig:SIMQSO-QSO}) and has a slightly larger KS distance than SFG and QSO (Table \ref{tab:experiments}).

In order to visualize the latent space, we show reconstruction images from a grid of points in the latent space (Fig. \ref{fig:re2d_map}). In each panel, the central image is the reconstruction for the average point in the latent space distribution, and from this central point the grid is created with two latent space vectors. The horizontal vectors point from the average position of the SFGs to the average position of the QSOs, while the two vertical vectors are random. The similarity of nearby images, clearly displayed in both panels of this figure, is one of the primary attributes of the VAE.

\subsection{Star-forming and quiescent galaxies} \label{sec:SFQ}

Compared to the previous section, the results of comparing SFG to QG are statistically more robust in the sense that the VAE is grouping both the general populations (Fig.~\ref{fig:KS}, panel $\#4$) and the outliers (p $\sim 10^{-46}$) more strongly, but statistically less robust in the sense that the outlying regions show strong correlations with the statistical quantities which we discussed in the previous section. This could mean either that for this experiment the VAE is systematically grouping images with similar properties together in the latent space, or that the VAE is using morphology, but that the morphology correlates with the astrophysical statistics in a systematic way (or of course, some combination of these two explanations).

Disclaimers aside, a visual inspection of Fig.~\ref{fig:gal_QG} reveals a clear morphological difference between the populations, namely that image in the SFG region tend to have red central components, possibly consistent with a bulge, whereas the images in the QG region do not tend to have this feature.

\subsection{Effect of mask size} \label{sec:mask}

\begin{figure}[ht!]
\centering
\includegraphics[width=0.5\textwidth]{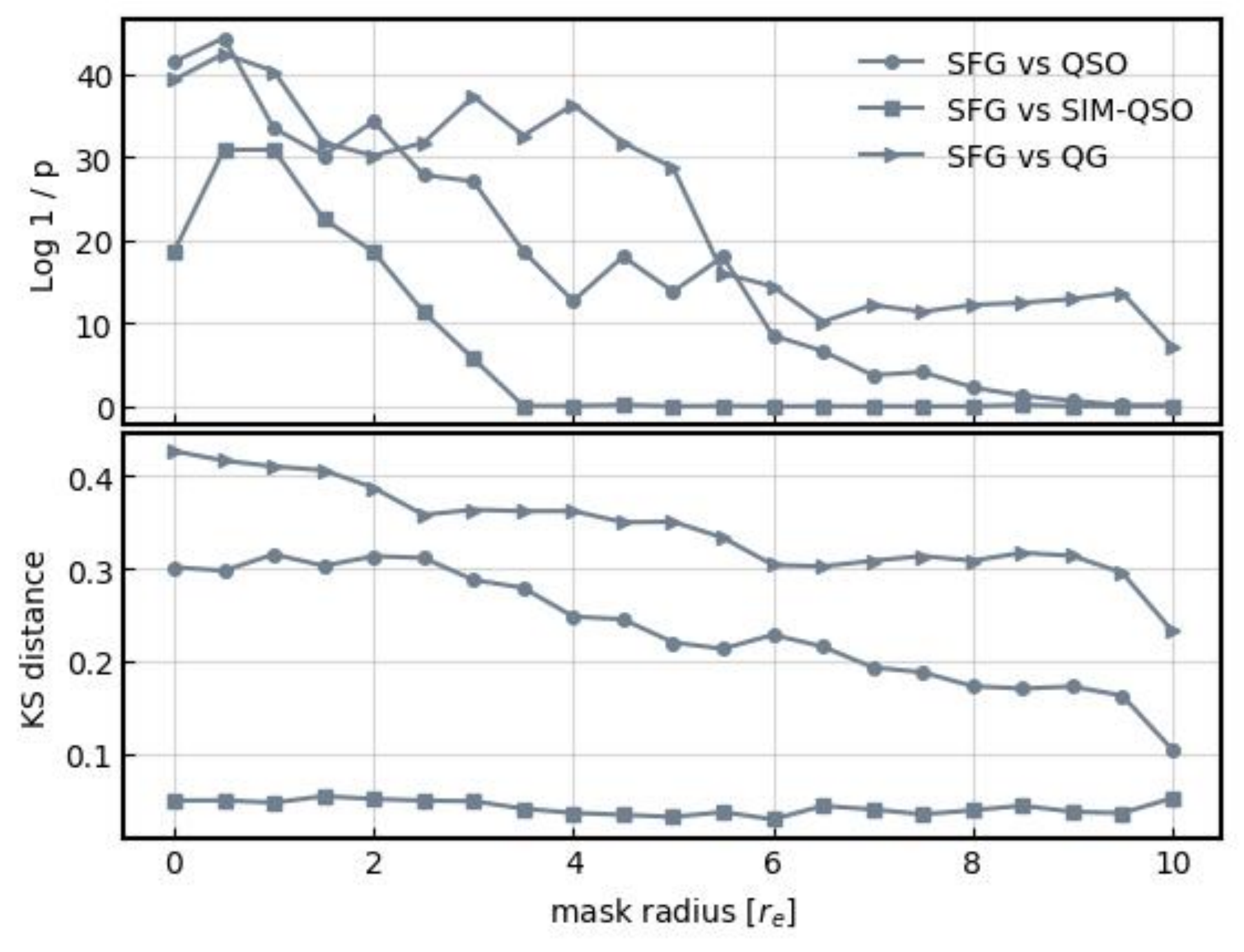}
\caption{Dependence on mask size of the probability from Sec. \ref{sec:latent} (upper panel) and the difference in KS statistics (lower panel) for the first three rows of Table \ref{tab:experiments}. \label{fig:mask_p}}
\end{figure}

In order to show at what radius the VAE can no longer detect the lingering effects of poor fits to the PSF (Sec.~\ref{sec:SFAGN}), we run the VAE on images with a circular mask applied to the center of the image. We do this for several VAEs of increasing mask size (Fig.~\ref{fig:mask_p}). We find that the mask completely obscures this effect for SIM-QSOs (compared to SFGs) at a radius of $3.5\times r_e$ when the probability reaches unity, indicating a randomly distributed latent space. This radius is larger than the expected size of the PSF ($\sim$ $r_e$), and demonstrates that a poor fit can effect a region of the image larger than that covered by the PSF. We conclude that the presence of an artifact from the PSF subtraction can be found by the VAE out to this radius, but we emphasize that this only effects a few of the SIM-QSO images as the corresponding KS test (bottom panel) is always close to zero. 

\begin{figure}[ht!]
\centering
\includegraphics[width=0.5\textwidth]{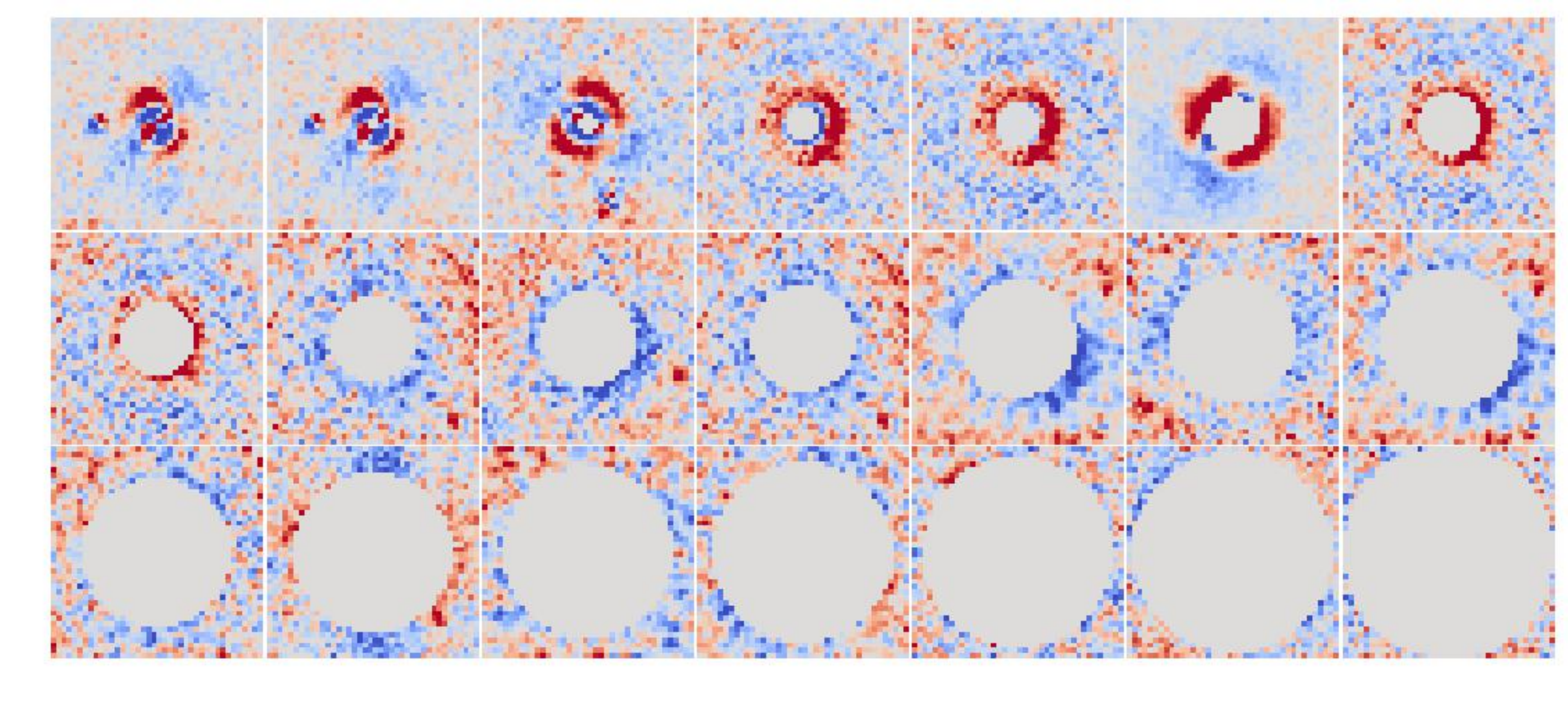}
\caption{The image with the highest QSO fraction in the latent space, for the VAEs trained on SFG/QSO in Fig.~\ref{fig:mask_p} with different sized masks.  \label{fig:mask_im}}
\end{figure}

We also conduct this same test for the other two experiments, and the image with the highest QSO fraction for the SFG vs QSO experiment is shown in Fig.~\ref{fig:mask_im}. For larger mask sizes, the VAE seems to be using almost exclusively the region directly surrounding the mask (red/blue rings around the mask). One explanation for this is that the VAE is good at recognizing edges. For the SFG vs QSO experiment, the probability is mostly increasing as a function of mask size, until it eventually reaches unity (that is, a random latent space). The KS statistic is level until the mask reaches $2.5\times r_e$, after which it decreases.

\section{Summary and concluding remarks} \label{sec:discussion}

We have tested whether there are structural differences between galaxies hosting a quasar and a matched control sample of star-forming galaxies. For this exercise, we have removed the quasar emission, a smooth model of the host galaxy emission, and nearby neighbors based on a 2D decomposition of optical images of SDSS quasars from the Subaru HSC-SSP survey. These residual images allow us to investigate the level of (sub)-structure internal (and external) to quasar hosts. From visual inspection, these residual images can exhibit familiar features such as spiral arms, arcs, bars, rings, and clumps. It is these features that may further our understanding of how gas is driven to the nuclear region to provide fuel for accretion onto SMBHs through a secular means. The importance of such studies are elevated by the fact that major mergers are a minor contributor to black hole growth for this quasar sample \citep[e.g.][]{tang2023morphological}.


With the use of a variational auto-encoder, we constructed a lower dimensional latent space into which these residual images were placed such that nearby images in the latent space have similar visual, and hence morphological, properties. We then analyzed this distribution to see if various populations were differentiated globally and within specific regions of the latent space.

We found that there are significant differences in the distribution of QSOs and SFGs within the latent space. The fraction of a population, either QSOs or SFGs, within the 200 nearest neighbors is dissimilar between the two (Fig.~\ref{fig:KS}, panel $\#2$). As a check, we ran the same analysis on a sample of simulated QSOs and their star-forming hosts to ensure that the latent space distributions were not due to an artifact induced by the removal of the PSF. The fraction of nearest neighbors for the simulated QSO sample is consistent with the SFG population. For the real QSOs, we determined that the differences are at least in part morphological by performing a visual inspection of the regions dominated by one population or the other (Figs. \ref{fig:gal_QSO}, \ref{fig:SIMQSO-QSO}; Appendix A). In the case of SFG vs QSO, the QSO morphology appears to correspond to more prominent features such as rings or spiral arms that may indicate a connection between secular processes and quasar activity. There is also the possibility of remnant features (i.e., tidal tails) from ongoing or past merger events contributing to the differences seen here which may be related to the fact that the quasars hosts are, on average, more compact than star-forming galaxies 
\citep{Li2021ApJ...918...22L} and more asymmetric for the more luminous cases (Tang et al. submitted). In addition and as expected, we found that QGs and SFGs are differentiated both globally and locally.

In conclusion, the VAE finds a difference between the SFG and QSO populations. The difference appears to be morphological and it does not appear to be statistical, but we cannot be sure whether the VAE is in fact relying on morphology. In any case, we demonstrate the use of the VAE as a powerful tool for the image analysis of quasar hosts and their internal structure. However, the existence of many hyper-parameters and steps in the process of image preparation demand further exploration and optimization of this tool. Larger data-sets (i.e., Rubin/LSST) will greatly improve the statistical robustness of our results as well as making more clear exactly which morphological features the VAE is relying on.

 \section*{Acknowledgements}

This work was supported by World Premier International Research Center Initiative (WPI), MEXT, Japan. J.S. is supported by JSPS KAKENHI (grant Nos. JP18H01251 and JP22H01262) and the World Premier International Research Center Initiative (WPI), MEXT, Japan. TH acknowledges funding from JSPS KAKENHI Grant Number 20K14464.



\bibliography{bib}{}
\bibliographystyle{aasjournal}


\section{Appendix A: Gallery of outlying regions}

\begin{figure*}[ht!]
\centering
\includegraphics[width=0.9\textwidth,height=0.45\textwidth]{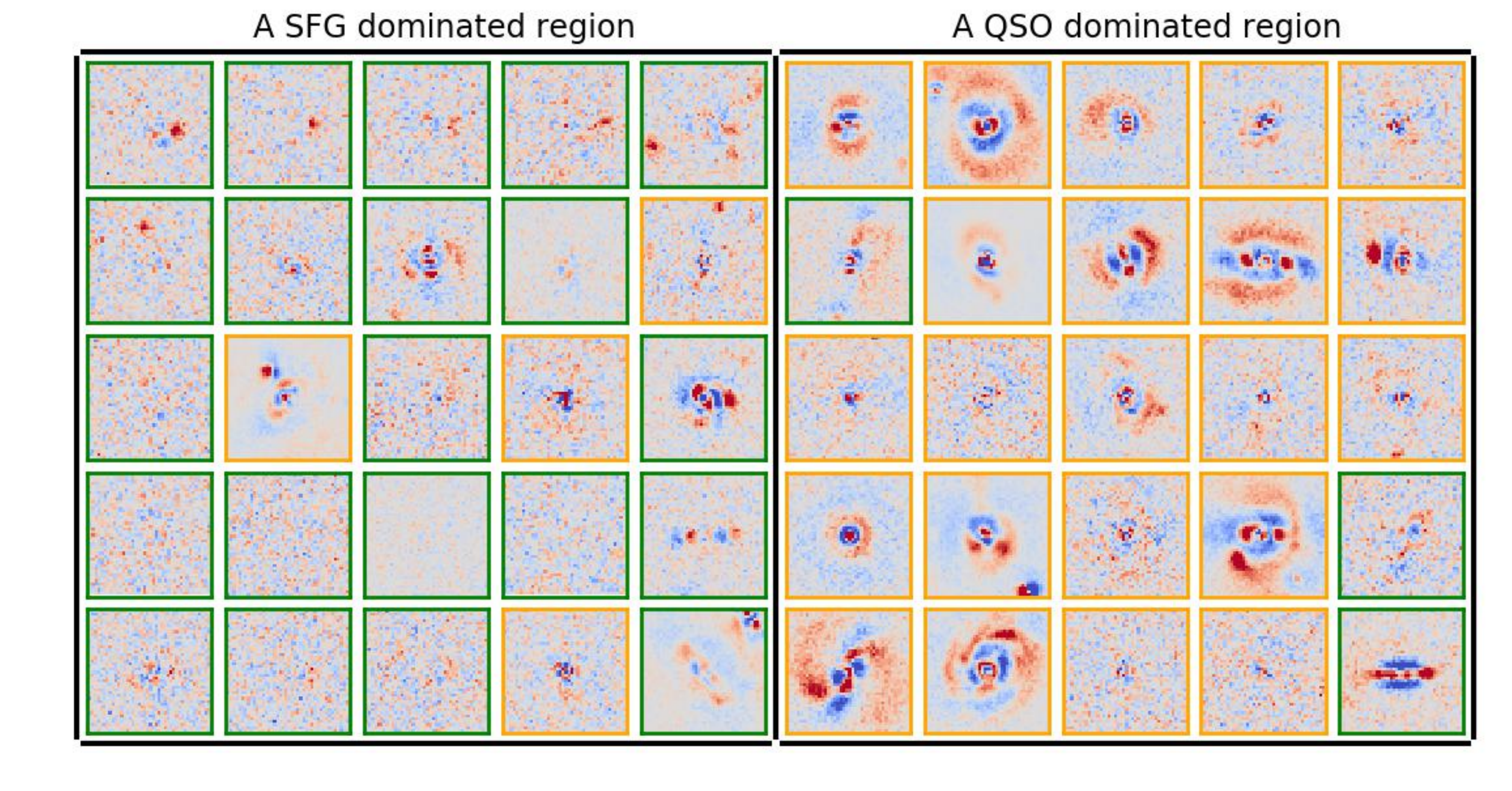}
\caption{ Same as Fig.~\ref{fig:gal_QSO} but for a different outlying region. The QSO dominated region shows evidence of structures on even larger scales than Fig. \ref{fig:gal_QSO}, as well as possible evidence of bars. \label{fig:c11}}
\end{figure*}

\begin{figure*}[ht!]
\centering
\includegraphics[width=0.9\textwidth,height=0.45\textwidth]{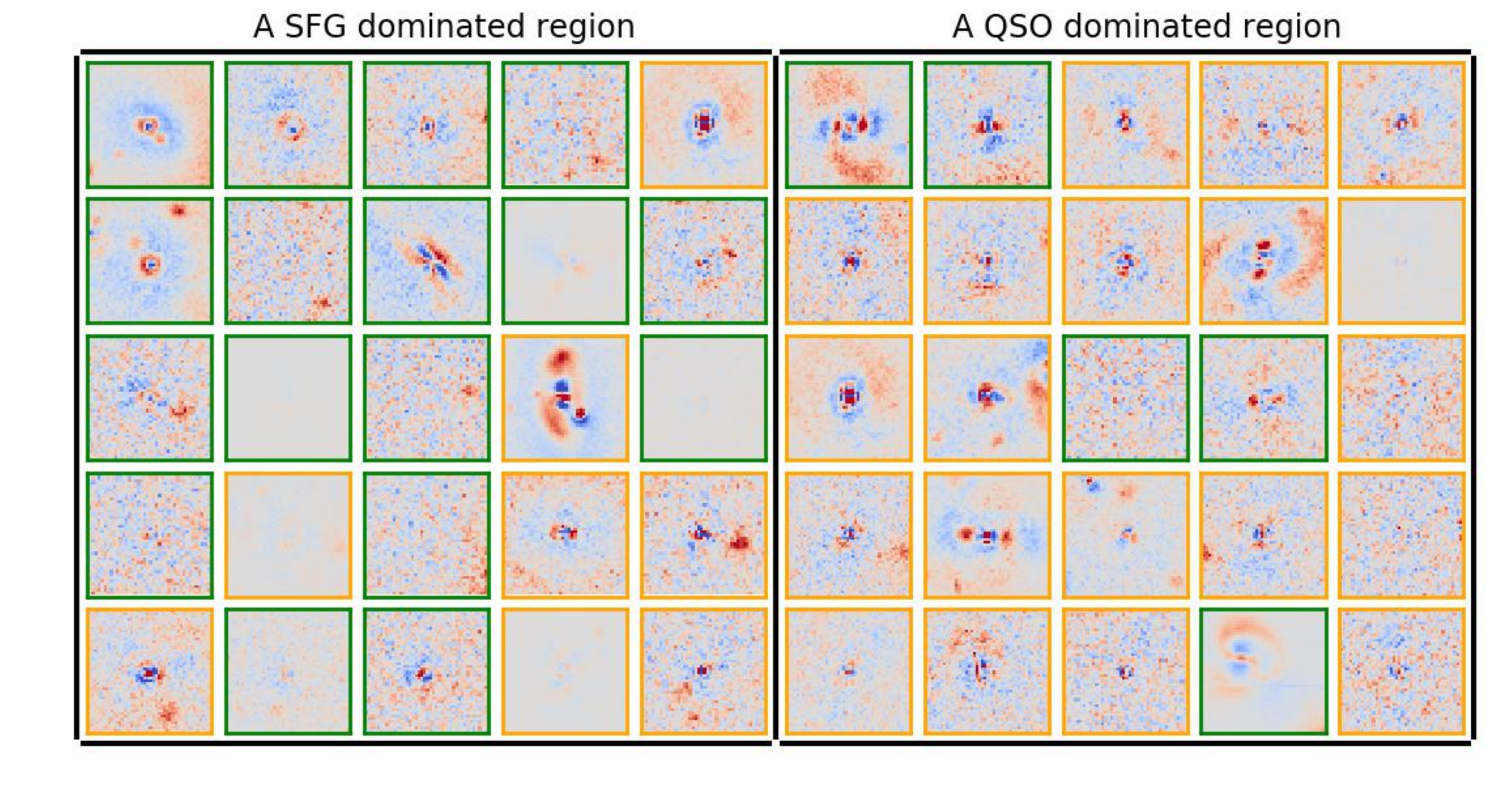}
\caption{ Same as Fig.~\ref{fig:gal_QSO} but for a different outlying region. The QSO dominated region shows limited evidence of very large scale spiral arms.\label{fig:c14}}
\end{figure*}

\begin{figure*}[ht!]
\centering
\includegraphics[width=0.9\textwidth,height=0.45\textwidth]{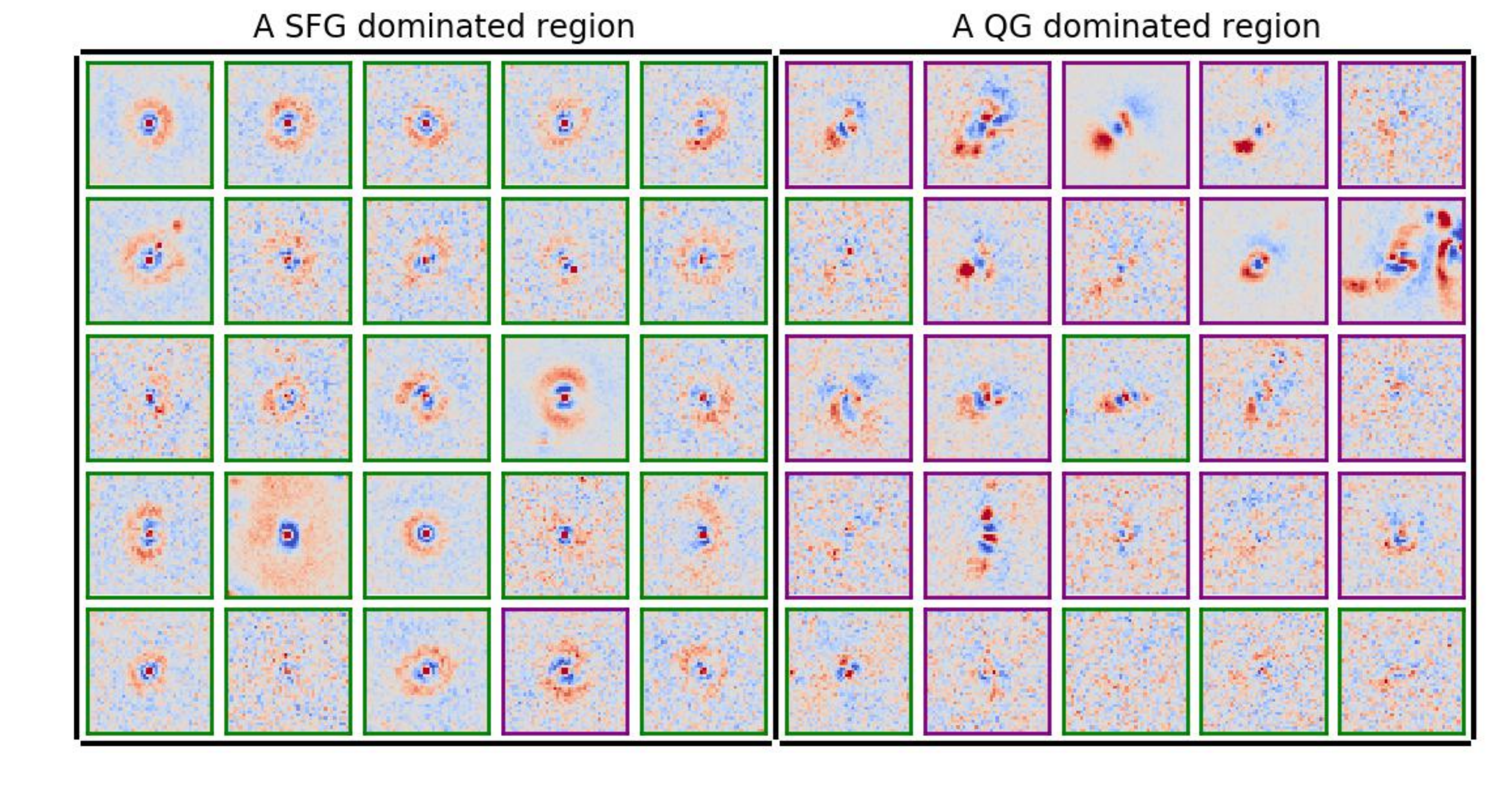}
\caption{ Same as Fig.~\ref{fig:gal_QG} but for a different outlying region. The SFG dominated region shows evidence of symmetric, ring-like structure. \label{fig:c01}}
\end{figure*}

\begin{figure*}[ht!]
\centering
\includegraphics[width=0.9\textwidth,height=0.45\textwidth]{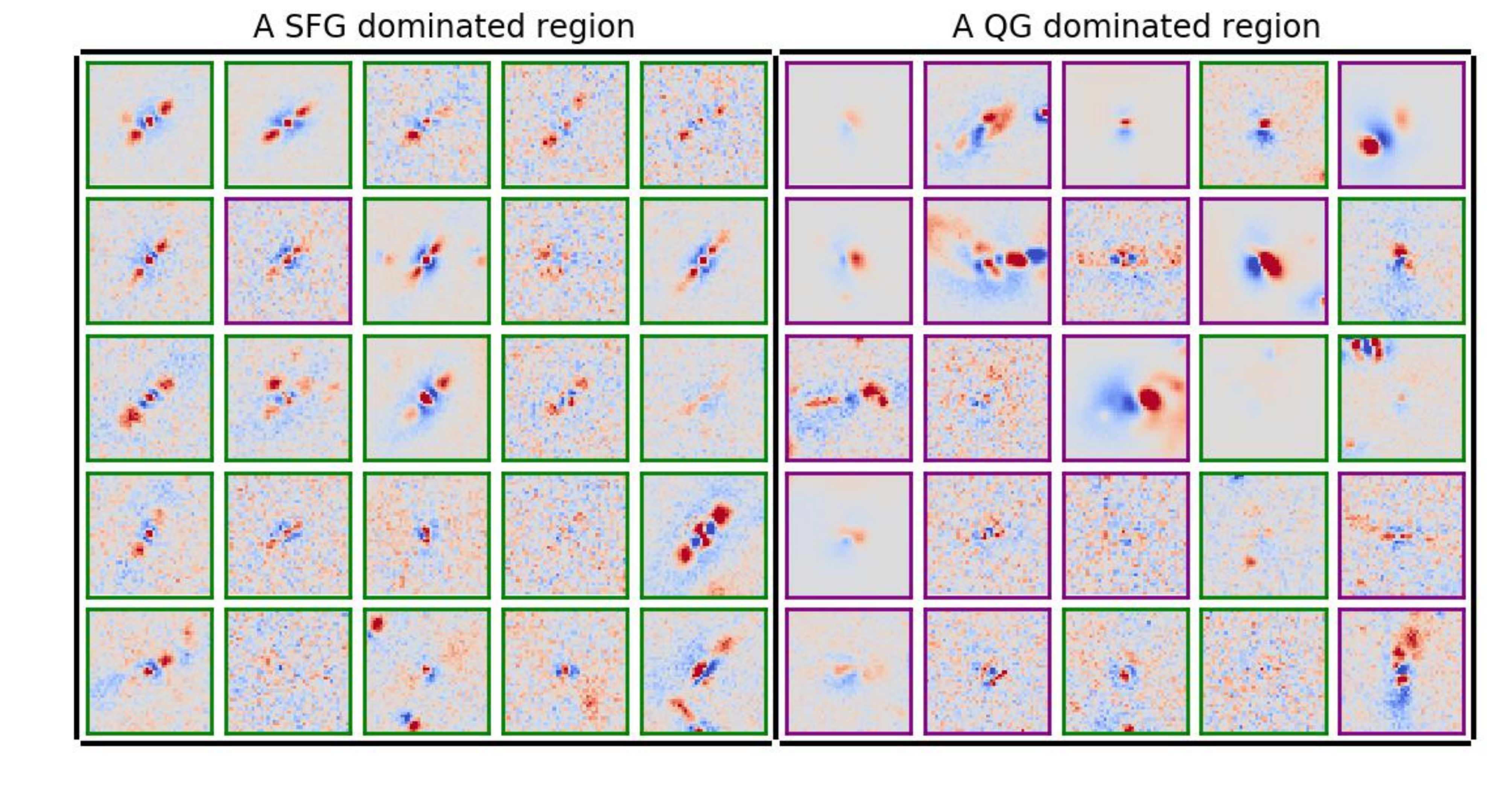}
\caption{ Same as Fig.~\ref{fig:gal_QG} but for a different outlying region. The SFG dominated region shows evidence of edge on disks.\label{fig:c03}}
\end{figure*}

\begin{figure*}[ht!]
\centering
\includegraphics[width=0.9\textwidth,height=0.45\textwidth]{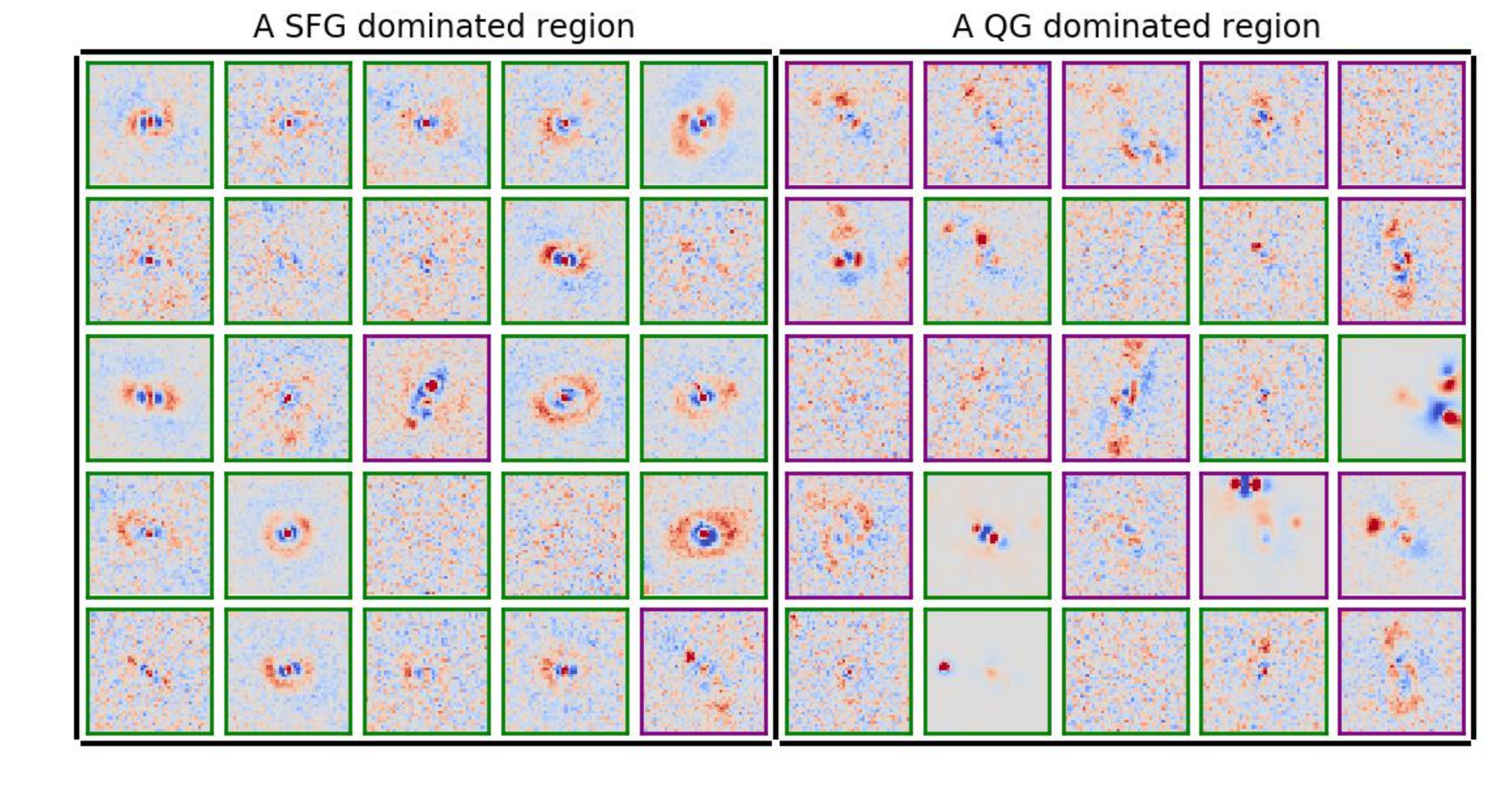}
\caption{ Same as Fig.~\ref{fig:gal_QG} but for a different outlying region. The SFG dominated region shows evidence of ring-like structure.\label{fig:c05}}
\end{figure*}

\begin{figure*}[ht!]
\centering
\includegraphics[width=0.9\textwidth,height=0.45\textwidth]{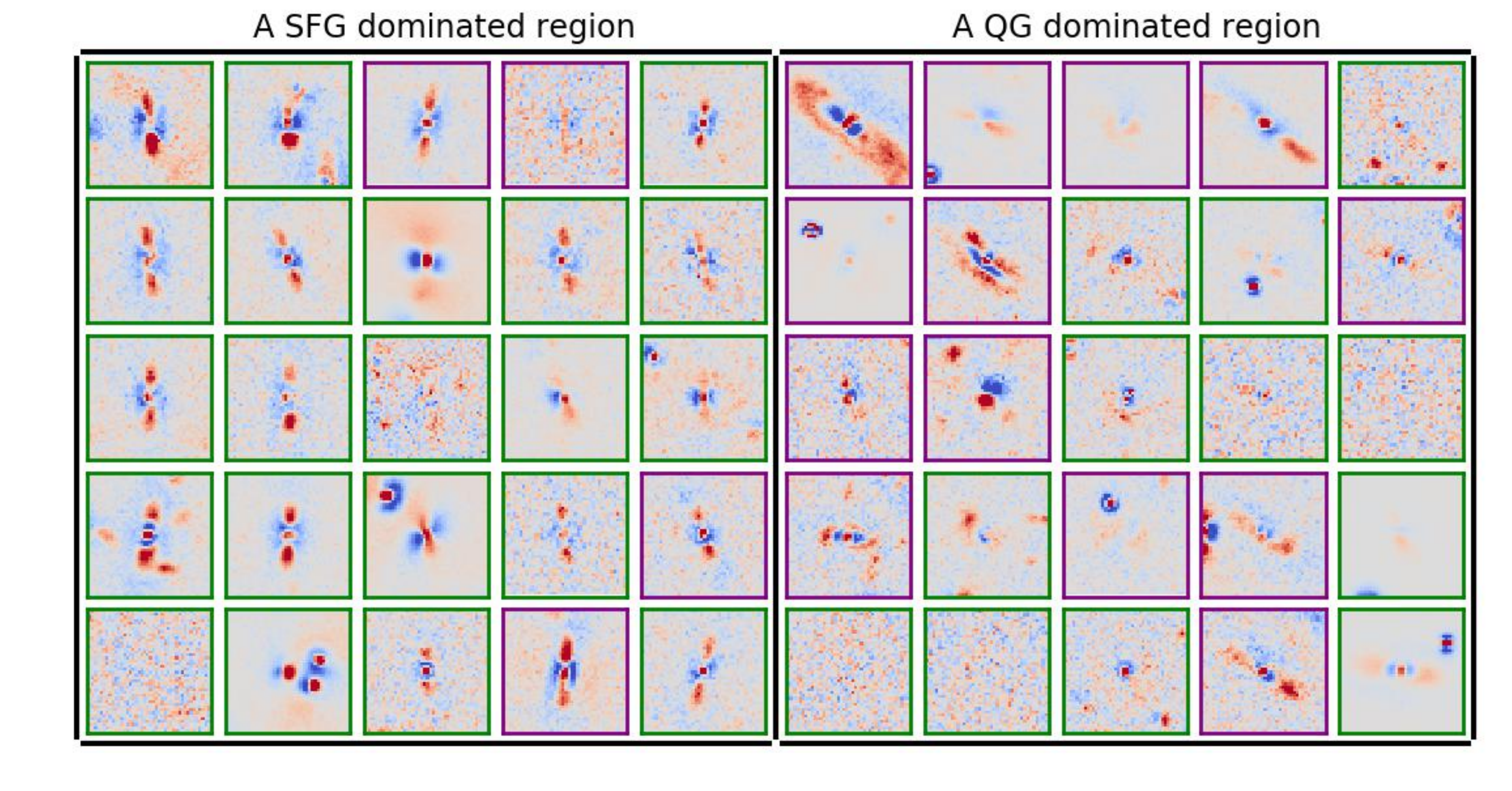}
\caption{ Same as Fig.~\ref{fig:gal_QG} but for a different outlying region. The SFG dominated region shows evidence of edge on disks. The QG region shows large elliptical structure. \label{fig:c06}}
\end{figure*}

In the below images (Figs. \ref{fig:c11}, \ref{fig:c14}, \ref{fig:c01}, \ref{fig:c03}, \ref{fig:c05}, \ref{fig:c06}), we show unique outlying regions which seem to show morphological differences between SFG and QSO and between SFG and QG. Note that there are some outlying regions which do not show clear morphological differences, but they are in the minority.

\end{document}